\documentclass[11pt]{article}

\usepackage{amsmath, amssymb, amsbsy, braket, bm, dsfont}
\usepackage{graphicx}
\usepackage{subfigure}		% This is supposed to be obsolete, superseded by subfig
\usepackage{epstopdf}
\usepackage[breaklinks]{hyperref}

\newcommand{\phd}{ {\vphantom{\dag}} }		% phantom dagger

\newcommand{\dual}{ {\mathcal{D}} }
\newcommand{\hc}{ {\mathfrak{h.c.}} }
\newcommand{\ketbra}[2]{ {\ket{#1} \! \bra{#2}} }

\newcommand{\Zth}{\mathbb{Z}_3}
\newcommand{\zbar}{\overline{z}}
\newcommand{\Xbar}{{\mkern3.5mu\overline{\mkern-3.5mu X \mkern-1mu}\mkern1mu}}
\newcommand{\Tbar}{{\mkern1.5mu\overline{\mkern-1.5mu T \mkern-0.5mu}\mkern0.5mu}}
\newcommand{\vv}[2]{\Phi_{#1\overline{#2}}}

\usepackage[usenames,dvipsnames]{color}			% Only used for author comments
\definecolor{purple}{rgb}{0.5,0,0.5}
\definecolor{dkgreen}{rgb}{0,0.5,0}

\newcommand{\paul}[1]{ { \color{blue} \footnotesize (\textsf{PF}) \textsf{\textsl{#1}} } }
\newcommand{\comment}[1]{}

\begin{document}

\title{Parafermionic conformal field theory on the lattice}
\author{Roger S. K. Mong$^1$, David J. Clarke$^1$, Jason Alicea$^1$, Netanel H. Lindner$^{1,2}$, Paul Fendley$^{3}$
\\	${}^1$ Department of Physics and Institute for Quantum Information and Matter, \\California Institute of Technology, Pasadena, CA 91125, USA
\\	${}^2$ Department of Physics, Technion, 32000 Haifa, Israel
\\	${}^3$ Department of Physics, University of Virginia, Charlottesville, VA 22904-4714 USA
}

\date{\today}

\maketitle

\begin{abstract}
Finding the precise correspondence between lattice operators and the continuum fields that describe their long-distance properties is a largely open problem for strongly interacting critical points.
Here we solve this problem essentially completely in the case of the three-state Potts model, which exhibits a phase transition described by a strongly interacting `parafermion' conformal field theory.
Using symmetry arguments, insights from integrability, and extensive simulations, we construct lattice analogues of nearly all the relevant and marginal physical fields governing this transition. This construction includes chiral fields such as the parafermion.
Along the way we also clarify the structure of operator product expansions between order and disorder fields, which we confirm numerically.
Our results both suggest a systematic methodology for attacking non-free field theories on the lattice and find broader applications in the pursuit of exotic topologically ordered phases of matter.

\end{abstract}

%%%%%%%%%%%%%%%%%%%%%%%%%%%%%%%%%%%%%%%%%%%%%%%%%%%%%%%%%%%%%%%%%%%%%%%%%%%%%%%%
\section{Introduction}

A major triumph of theoretical physics was the precise understanding, via the renormalization group, of how lattice models of statistical mechanics are described by continuum field theories at and near critical points.
In 1+1-dimensional quantum critical points with linear dispersion and two-dimensional classical systems with rotational invariance, all fields and their exact scaling dimensions often can be identified precisely by using conformal field theory (CFT) \cite{BPZ}.
Once one identifies which particular CFT describes a lattice model of interest, it is then possible to utilize powerful non-perturbative techniques to do many exact computations.
Moreover, certain models---called \emph{integrable}---allow exact computations to be performed even away from criticality \cite{Baxbook}.

Despite the many successes of this approach, the connection between the original lattice formulation of a theory and its continuum counterpart is often difficult to make precise.
One question that frequently arises is the following: Given some microscopic operator $\hat{O}$, what is its expansion in terms of continuum fields at criticality?
When a CFT describes the continuum limit, we can turn the question on its head and ask: Which combination of lattice operators yields a \emph{particular} continuum field as the lowest-scaling-dimension component in such an expansion?
The former admits a unique answer, but the latter does not.

These questions are essential to understand for many applications involving critical phenomena---\emph{e.g.}, extracting correlation functions of physical quantities or perturbing critical points with specific microscopic operators.
Yet the answers are surprisingly incomplete given the vast body of literature on critical systems.
Resolving the connection between lattice operators and \emph{chiral} continuum fields has proven especially difficult.  %It is especially tricky to find precisely which lattice operator corresponds to a specific \emph{chiral} field in the continuum limit.
Here the additional challenge derives from the fact that a chiral operator on the lattice is, by definition, non-local.

One reason chiral operators are particularly interesting is because of the many formal connections between chiral fields in CFT and anyons in topologically
ordered systems.
Although it may naively seem as if studying a lattice model is not necessary to analyze an elegant continuum theory, one central point of this paper is that this exercise not only results in a great deal of intuition into the problem, but also uncovers deep results.
For example, the lattice analogues of certain local fields in minimal CFTs are given in terms of single-site operators using the modular $S$ matrix of topological field theory \cite{PasquierOp}.
In fact, this analysis of Pasquier's resulted in his discovery of a special case of the Verlinde formula \cite{Verlinde} before Verlinde!
%Here we will further explore such connections.  \jason{Do we still do this?  Or is this stuff now designated for the sequel to the sequel?}

The Ising model comprises one of the few examples where the lattice/continuum correspondence is essentially completely understood.
Here there are only two non-trivial relevant local operators, the spin and energy fields, which respectively possess scaling dimensions $1/8$ and $1$.
Symmetry alone allows one to relate their `ultraviolet' and `infrared' manifestations.
For example, the spin field is the most relevant operator that is odd under the symmetry corresponding to a global spin flip.
Thus the operator measuring the spin at a lattice site has the spin field as its leading contribution in the continuum limit.
The lattice analogue of the energy or ``thermal'' field is found simply by noticing that perturbing the 2d classical Ising model off the critical point by changing the temperature preserves the spin-flip symmetry.
The only symmetry-preserving relevant operator is the energy field, so the lattice expression can be extracted directly from the action (or Hamiltonian in the quantum spin-chain case).
The most interesting chiral operators in the Ising model, the left- and right-moving components $\bar\psi$ and $\psi$ of the free-fermion field, are also well understood.
These non-local operators are obtained by the so-called Jordan-Wigner mapping \cite{SML}, and can be elegantly understood as a product of spin and disorder operators \cite{KadanoffCeva}.
A simple consistency check on these relations follows from the fact that the energy field is the product $\psi\bar\psi$.

Unfortunately, analogous results are not so simple to obtain for more general models, even using very sophisticated techniques.
Great progress was made in the '70s using the Coulomb-gas approach pioneered by Kadanoff and others, where critical properties of a wide variety of classical lattice models were argued to be identical to those of a free boson, and then various heuristic arguments were used to identify the exact scaling dimensions of some operators \cite{Nienhuis}.
With the advent of conformal field theory in the '80s, these results were adapted and generalized.
An example close to hand here is the antiferromagnetic three-state Potts model \cite{Delfino}.
The ferromagnetic 3-state Potts model studied here also has been treated by this approach, but the Coulomb-gas results are even more heuristic, since the free-boson theory needs to be modified by including a charge at infinity \cite{denNijs,Dotsenko}.
Thus while this approach has yielded many valuable results, it typically is somewhat \emph{ad hoc}, and moreover rarely yields chiral operators. 

Another reason for revisiting the lattice/continuum correspondence arises from recent work in the mathematical physics/probability community. In the context of two-dimensional classical lattice models, lattice chiral operators are known as {\em discrete holomorphic} operators. One reason for mathematicians' interest is the potential that these can be used to rigorously prove that a given lattice model turns into a particular CFT in the continuum limit, a strategy successfully used in the Ising case \cite{Smirnovreview}. Another reason is that demanding an operator be discrete holomorphic in many cases provides a simple way of finding integrable models \cite{Cardydiscrete}. In fact, using considerations from topological field theory it is possible to find a general method for constructing discrete holomorphic operators of this type \cite{Pdiscrete}.

It turns out, however, that such definitions are typically not sufficient to fix lattice operators precisely.
Theories more complicated than Ising or a free boson admit multiple operators that behave similarly under discrete rotations; in CFT language such operators possess conformal spins differing by integers.
A given lattice operator will then represent some linear combination of these continuum operators, and it is not \emph{a priori} obvious how to separate the constituent pieces.
In the Ising case, the fact that the chiral fermions are non-interacting makes it possible to find additional constraints sufficient to determine the precise connection between lattice and continuum. In interacting cases, it has not been so. For example, the equations usually solved for discrete holomorphicity in classical models typically amount to lattice versions of only half the Cauchy-Riemann equations.

One of our main results is showing how in a \emph{non-free} field theory we can overcome this obstacle and precisely identify chiral lattice operators. The example is the famed three-state Potts model---which provides a very natural generalization of the Ising model in some respects, but is strongly interacting.
The three-state Potts model generalizes the Ising model by replacing the variable on each site by a three-state ``spin''.
With ferromagnetic nearest-neighbor interactions, the model is ordered at low temperatures and remains disordered at high temperatures, as with Ising.
%The Hamiltonian is given by
%\begin{align}
%	H = -J \sum_{a} \big( \hat\sigma^\dag_{a+1} \hat\sigma_{a}^\phd + \hat\sigma^\dag_{a} \hat\sigma_{a+1}^\phd \big)
%		- f \sum_{a} \big( \hat\tau^\dag_a + \hat\tau_a^\phd \big) \ ,
%		\label{Hpotts}
%\end{align}
%where $\hat\sigma_a$ and $\hat\tau_a$ denote operators that act on site $a$ and satisfy $\hat\sigma_a^3 = \hat\tau_a^3 = 1$ and the commutation relation $\hat\sigma_a \hat\tau_a = e^{2\pi i/3} \hat\tau_a \hat\sigma_a$.
%One can alternatively express $H$
A non-trivial critical point separates the two phases,  and the ${\mathbb Z}_3$ ``parafermion'' CFT describing the continuum limit is well understood \cite{FZ2}. The naming arises because the CFT includes a chiral parafermion field -- a ${\mathbb Z}_3$ analogue of the chiral fermion in Ising, with similar algebraic structure but without the feature that makes the model easily solvable. Likewise, the associated quantum lattice Hamiltonian, given in (\ref{eq:H}) below, can be rewritten in terms of lattice {parafermion} operators \cite{FK}.

As the lattice parafermion operator and CFT parafermion field share common symmetry properties, it is natural to expect that taking the continuum limit of the former recovers the latter.  We show, however, that the actual correspondence is more subtle.
On symmetry grounds one can not exclude the possibility that another continuum operator will also appear in an expansion of the lattice parafermion \cite{Rajabpour}; see Sec.~\ref{sec:chiral} for a detailed discussion.
In fact, this ``correction'' field turns out to exhibit a smaller scaling dimension than the parafermion field and thus, surprisingly, provides the dominant contribution to the expansion!
We construct a precise linear combination of lattice parafermion operators for which this contribution cancels, leaving the chiral parafermion field as the leading piece.  Arguments invoking symmetry and integrability allow us to similarly infer the lattice analogues of many other continuum fields. %, further solidifying the connection between the lattice model and the parafermion CFT.  % We derive the precise linear combinations relating various lattice operators and continuum fields (including the parafermions) using symmetry-based arguments.
We confirm these identifications by applying density-matrix renormalization group (DMRG) simulations to numerically compute two-point functions and hence the scaling dimensions of these lattice operators.  The DMRG method is particularly well-suited to this problem since it enables essentially exact computations for ground-state properties of one-dimensional quantum systems; in all cases we obtain perfect agreement with expectations based on our CFT-field correspondence.

% The methods we introduce are expected to enjoy broad applicability in various other contexts as well.  %Some of the relations derived here first appeared in Ref.~\cite{Mong}.

These findings further solidify the connection between the lattice model in Eq.~\eqref{eq:H} and the parafermion CFT.  Numerous other implications, however, also follow.  %The results discussed here have implications for physics beyond that exhibited by the Hamiltonian in Eq.~\eqref{Hpotts}.
On a formal level, our analysis clarifies the structure of operator product expansions in the CFT---particularly those that involve the parafermion fields, where in Sec.~\ref{sec:OPE} we identify an omission in previous results.  Furthermore, just as the Ising model has been instrumental in constructing phases that support Ising non-Abelian anyons, so too has the three-state Potts model been central to accessing phases with more exotic anyonic content.  Read-Rezayi quantum Hall phases \cite{RR} provide a classic example where parafermions play a key role.  More recently, Teo and Kane introduced a coupled-chain construction of Read-Rezayi states by hybridizing counterpropagating parafermion fields from adjacent critical chains.  Inspired by their work, the present authors and other collaborators \cite{Mong} introduced a superconducting analogue of the ($\Zth$) Read-Rezayi phase by weakly coupling critical Potts chains. Such a system can be `engineered' in Abelian quantum Hall/superconductor hybrids that localize lattice parafermion zero modes \cite{para, Clarke, Lindner, Cheng, BarkQi,Vaezi}.  This phase is remarkable in that it supports Fibonacci anyons---which possess universal braid statistics---yet is built from well-understood Abelian phases of matter.  Some of the relations derived here were first quoted in Ref.~\cite{Mong}; indeed, the expansion of the lattice parafermion operators that we elucidate below proved key to the entire analysis.  The present manuscript thus puts these earlier results on firmer footing and greatly expands them.  We expect the methodology pursued here to enable similar progress in other lattice systems, possibly paving the way to constructions of still more exotic two-dimensional phases of matter from critical chains.

\comment{
Perturbative methods such as the epsilon expansion have resulted in a wealth of information about critical behavior.
Many profound open questions remain though... non-perturbative approaches have long provided useful tools. These typically are in spatial dimensions lower than three, and so were mainly valuable for theoretical insight. Now due to experimental advances, they are also valuable tools for probing some of the most interesting systems in condensed-matter physics.
}

%%%%%%%%%%%%%%%%%%%%%%%%%%%%%%%%%%%%%%%%%%%%%%%%%%%%%%%%%%%%%%%%%%%%%%%%%%%%%%%%
\section{The three-state Potts model and its symmetries}
\label{sec:symmetries}

An obvious way to generalize the two-dimensional classical Ising model is to replace the variable on each site by a $q$-state ``spin''. When the interactions are nearest neighbor and only depend on whether the adjacent spins are the same or different, this is known as the $q$-state Potts model. In an isotropic system, there is then just one coupling, which can be taken to be the temperature.
With ferromagnetic interactions, the model orders at low temperatures and remains disordered at high temperatures, as with Ising.
Also as with Ising, there is a duality symmetry exchanging high and low temperatures, and the phase transition occurs at the self-dual point.  As opposed to Ising, however, this phase transition is first order for $q>4$ \cite{TL}. There does occur a self-dual critical point in a $q$-state model when the $S_q$ symmetry permuting the spins is broken down to $\mathbb{Z}_q$. This ``parafermion'' \cite{FK} critical point is integrable \cite{FZ1}; conserved charges have been computed explicitly \cite{paracharges}.
% However, moving away from the critical temperature while preserving the $\mathbb{Z}_q$ symmetry breaks the integrability except for the Ising case $q=2$, although the field theory describing the scaling limit near the critical point does remain integrable \cite{ZamoPotts,Fateev}.
Much of what we say in the following has an analogue for general $q$, but there the fine tuning necessary to extract the physics of interest is considerable. We therefore will confine our analysis to the three-state Potts model. %In a companion paper ???, we study one particularly interesting (integrable) perturbation.

\subsection{The Hamiltonian and the spin operators}

It is both intuitively and technically convenient to study the physics of the three-state Potts model by taking an anisotropic limit where the system can be described by a quantum Hamiltonian.
Taking this approach also has the advantage of making direct contact with the physics discussed in Ref.~\cite{Mong}.
The Hilbert space for an $L$-site chain is $(\mathbb{C}^3)^{\otimes L}$, i.e., a three-state system at each lattice site.
The $S_3$ symmetry permutes the three orthogonal basis states on each site.
The Hamiltonian for the three-state Potts quantum chain is
\begin{align}
	H = -J \sum_{a} \big( \hat\sigma^\dag_{a+1} \hat\sigma_{a}^\phd + \hat\sigma^\dag_{a} \hat\sigma_{a+1}^\phd \big)
		- f \sum_{a} \big( \hat\tau^\dag_a + \hat\tau_a^\phd \big)\ .
	%	- 	J \big( \sigma_1^\dag \sigma_N^\phd + \sigma^\dag_N \sigma_1^\phd \big)	.
	\label{eq:H}
\end{align}
Throughout we assume $J,f>0$.
The operator $\hat\tau_{a}$ shifts the spin on site $a \in \mathbb{Z}$, while $\hat\sigma_a$ measures its value.
Precisely, denoting the three states by $A$, $B$ and $C$, the operators on a particular site can be written as
\begin{align}
	\hat\sigma  &=  \ketbra{A}{A} + \omega\ketbra{B}{B} + \omega^2\ketbra{C}{C} =  \begin{pmatrix} 1&& \\ &\omega& \\ &&\omega^2 \end{pmatrix} ,
\\	\hat\tau  &=  \ketbra{B}{A} + \ketbra{C}{B} + \ketbra{A}{C} = \begin{pmatrix} &&1 \\ 1&& \\ &1& \end{pmatrix} .
	\label{eq:matrixrep}
\end{align}
where $\omega=e^{2\pi i/3}$.
These operators obey the algebra
\begin{equation}
		\hat\sigma_a^3 = 1\ ,
	\qquad \hat\tau_a^3 = 1\ ,
	\qquad \hat\sigma_a \hat\tau_a = \omega \hat\tau_a \hat\sigma_a\ ,
	\qquad \sigma_a\tau_b = \tau_b\sigma_a \textrm{ for $a \neq b$}.
\end{equation}
For $J > f $ the ground state forms a ferromagnet that spontaneously breaks the $S_3$ symmetry,
% with $\hat\sigma$ operator,
while for $f > J$ a disordered paramagnetic phase arises.
At the phase transition point $J = f$ the system is critical; we defer discussion of its critical properties to the next section.

%\subsection{Symmetries of the Hamiltonian}

Equation \eqref{eq:H} exhibits a number of symmetries that play an important role throughout this paper.
On an infinite chain (or a chain with periodic boundary conditions), the Hamiltonian preserves simple translations that shift the $\hat\sigma_a, \hat\tau_a$ operators by one site.
The Hamiltonian's full $S_3$ permutation symmetry can be usefully decomposed into a $\Zth$ symmetry---which cyclically permutes $|A\rangle$, $|B\rangle$ and $|C\rangle$---and a unitary `charge conjugation' operation ${\cal C}$ that swaps $\ket{B} \leftrightarrow \ket{C}$.
The $\Zth$ symmetry is generated by
\begin{align}
	\mathcal{Q} = \prod_a \hat\tau_a^\dag
\end{align}
and transforms operators according to $\hat{O} \rightarrow \mathcal{Q} \hat{O} \mathcal{Q}^\dag$.  In particular, we have
\begin{align}
	\hat\sigma_a &\rightarrow \omega\hat\sigma_a ,
&	\hat\tau_a &\rightarrow \hat\tau_a .
\end{align}
One can therefore say that $\hat\sigma_a$ carries `charge' $Q_3 = 1$ under $\Zth$ whereas $\hat\tau_a$ is neutral.
Here we take $\omega^{Q_3}$ to be the eigenvalue under $\mathcal{Q}$, so thus $Q_3$ is defined modulo 3.
Charge conjugation ${\cal C}$ acts on operators $\hat{O}$ via
\begin{align}
	\mathcal{C}[\hat{O}] = \left(\prod_a\hat{C}_a\right) \hat{O} \left(\prod_a\hat{C}_a\right) .
\end{align}
where %$\hat{C}_a$ is defined such that
$\hat{C}_a \hat\sigma_a \hat{C}_a = \hat\sigma_a^\dag$, $\hat{C}_a \hat\tau_a \hat{C}_a = \hat\tau_a^\dag$ and $\hat{C}_a^2 = 1$.
%, i.e.\ on a single site
%\begin{align}
%	\hat{C} = \ketbra{A}{A} + \ketbra{B}{C} + \ketbra{C}{B} .
%\end{align}
Note that $\cal{C}$ swaps the sign of the $\Zth$ charge carried by $\hat\sigma$---hence the term `charge conjugation'.
The Hamiltonian is also invariant under both parity (spatial inversion) and time-reversal symmetry.
Parity takes site $a$ to site $-a$; that is, $\mathcal{P}[\hat\sigma_a] = \hat\sigma_{-a}$ and $\mathcal{P}[\hat\tau_a] = \hat\tau_{-a}$.
The time-reversal generator $\mathcal{T}$ is anti-unitary and conjugates $\hat\sigma_a$ ($\mathcal{T}[\hat\sigma_a] = \hat\sigma_a^\dagger$) but leaves $\hat\tau_a$ invariant ($\mathcal{T}[\hat\tau_a] = \hat\tau_a$).

\subsection{Dual variables}
\label{sec:dual}

While the preceding section enumerated the complete set of symmetries manifest in the three-state Potts Hamiltonian expressed in terms of the spin operators $\hat\sigma_a$ and $\hat\sigma_a^\dagger$, additional symmetries are revealed upon recasting the model in dual variables.
Namely, {\em disorder operators} \cite{KadanoffCeva} can be defined here using a generalization of the Kramers-Wannier duality of the Ising model.
In the quantum Hamiltonian limit they are defined as
\begin{align}
	\hat\mu_b  \equiv  \prod_{a < b} \hat\tau_a , \qquad	b \in \mathbb{Z}+\tfrac{1}{2} ,
	\label{eq:def_mu}
\end{align}
which live on \emph{bonds} between sites of the original lattice.  We use conventions where these bonds are labeled by half-integers---e.g., the bond between sites $a$ and $a+1$ is denoted by $b = a + \frac{1}{2}$.
The disorder operator $\hat\mu_b$ in effect adds a domain wall between sites $b-\frac{1}{2}$ and $b+\frac{1}{2}$ by cycling all spins to the left of bond $b$.
Conjugating the Hamiltonian by the disorder operator, $\hat \mu_b H \hat \mu_b^\dag$, leaves all terms invariant except for the ferromagnetic $J$ term that couples  sites $b\pm\frac{1}{2}$.
This suggests that the operator conjugate to $\hat\mu_b$ ought to be
\begin{align}
	\hat\nu_b  \equiv  \hat\sigma_{b-\frac12}^\dag \hat\sigma_{b+\frac12}^\phd .
	\label{eq:def_nu}
\end{align}
Indeed, the $\hat\mu$ and $\hat\nu$ operators defined as above obey exactly the same algebra as $\hat\sigma$ and $\hat\tau$:
\begin{align}
	\hat\mu_b^3 &= \hat\nu_b^3 = 1 ,
&	\hat\mu_a \hat\nu_b &= \begin{cases} \omega \hat\nu_b \hat\mu_a & a=b , \\ \hat\nu_b \hat\mu_a & a \neq b . \end{cases}
	\label{eq:munu}
\end{align}
In a very precise sense these operators are the duals of the original spin operators.

Remarkably, the Hamiltonian written in terms of $\hat\mu$, $\hat\nu$ retains the same form as Eq.~\eqref{eq:H},
\begin{align}
	H = - f \sum_{b \in \mathbb{Z}+\frac12} \big( \hat\mu^\dag_{b+1} \hat\mu_{b}^\phd + \hat\mu^\dag_{b} \hat\mu_{b+1}^\phd \big)  -J \sum_{b \in \mathbb{Z}+\frac12} \big( \hat\nu^\dag_b + \hat\nu_b^\phd \big)	 ,
\end{align}
with the roles of $J$ and $f$ swapped.
Hence we can define the \emph{duality} transformation
\begin{align}
	\dual[ \hat\sigma_a ] &= \hat\mu_{a+\frac12} , & \dual[ \hat\tau_a ] &= \hat\nu_{a+\frac12} ,
	\label{eq:dualityT}
\end{align}
which is an additional symmetry of the Hamiltonian at the critical point $J = f$.
Since the transformation takes $\hat\mu_a \rightarrow \hat\sigma_{a+\frac12}$ and $\hat\nu_a \rightarrow \hat\tau_{a+\frac12}$,
	two applications of the duality transformation yield a simple lattice translation by one site: $\dual^2[\hat{\sigma}_a] = \hat{\sigma}_{a+1}$ and $\dual^2[\hat{\tau}_a] = \hat{\tau}_{a+1}$.
In this representation one can also identify a $\Zth^\textrm{dual}$ symmetry (present even for $J \neq f$) generated by $\mathcal{Q}^\textrm{dual} = \prod_b \hat \nu_b^\dag$.
This symmetry acts on the dual operators according to
\begin{align}
	\hat\mu_b &\rightarrow \omega\hat\mu_b ,
%&	\hat\mu'_b &\rightarrow \omega\hat\mu'_b ,
&	\hat\nu_b &\rightarrow \hat\nu_b .
\end{align}
Hence $\hat\mu_b$ and $\hat\nu_b$ respectively possess charge $Q_3^{\rm dual} = 1$ and $Q_3^{\rm dual} = 0$ under $\Zth^\textrm{dual}$.
For an infinite chain, one can interpret this transformation as a phase applied to $\hat\tau_{\pm\infty}$ (since here $\mathcal{Q}^\textrm{dual} \sim \hat\sigma_{-\infty}\hat\sigma_{+\infty}^\dagger$).
For a periodic chain, the transformation can instead be thought of as acting on the space of boundary conditions.

Importantly, the duality mapping exploited above is not unique.  Instead of defining the disorder operator in terms of a string of $\hat\tau_a$'s emanating to $-\infty$, we can of course also choose the opposite convention with%Of course, instead of defining the string of the disorder operator out of $-\infty$, we may also choose the opposite convention with
\begin{align}
	\hat\mu'_b &= \prod_{a > b} \hat\tau_a^\dag.
	\label{eq:def_mu'}
\end{align}
This `primed' disorder operator creates a domain wall by cycling the spins to the \emph{right} of bond $b$, and relates to the disorder operator defined earlier via $\hat\mu_b' = \mathcal{Q} \hat\mu_b$.
%\paul{My suggestion is to not define primes for $\mu$, but rather just use the operator $Q$ (what I call $\omega^{\cal P}$ in my papers)}.
Despite the reversed string orientation in $\hat\mu_a'$, the operator $\hat\nu_a$ that winds the (dual) spin measured by $\hat\mu_a'$ is again defined as in Eq.~\eqref{eq:def_nu}, and the relations~\eqref{eq:munu} hold true with $\hat\mu$ replaced by $\hat\mu'$. %\footnote{We note the useful identity is $\mu_a' = \mathcal{Q} \mu_a$.}
Consequently there is an alternative duality transformation which satisfies
\begin{subequations}\begin{align}
	\dual'[\hat\sigma_a] &= \hat\mu^{\prime\dag}_{a-\frac12} ,
&	\dual'[\hat\tau_a] &= \hat\nu_{a-\frac12}^\dag ,
\\	\dual'[\hat\mu'_b] &= \hat\sigma_{b-\frac12}^\dag ,
&	\dual'[\hat\nu_b] &= \hat\tau_{b-\frac12}^\dag .
\end{align}\end{subequations}
This alternative duality is simply a mirror version of our earlier definition: $\mathcal{D}'[\cdot] = \mathcal{P}[\mathcal{D}[\mathcal{P}[\cdot]]]$.
Two applications thus again yields a simple lattice translation, i.e., $\dual^{\prime2}[\hat\sigma_a] = \hat\sigma_{a-1}$ and $\dual^{\prime2}[\hat\tau_a] = \hat\tau_{a-1}$.

Table~\ref{tab:OpSym} summarizes the symmetry properties of the original Potts operators, their duals, and the generators $\mathcal{Q}$, $\mathcal{Q}^{\rm dual}$ of $\Zth$, $\Zth^{\rm dual}$ transformations.
(Lattice translations are suppressed since they act trivially on all operators.)
Note that parity swaps the two types of disorder operators $\hat\mu_a$ and $\hat\mu_a'$.  This fact will become important in Sec.~\ref{sec:chiral}.

\begin{table}
	\begin{align*} \renewcommand*{\arraystretch}{1.1} \begin{array}{c|cc|cccc|cc}
		\textrm{Operator} & Q_3 & Q_3^\textrm{dual}
		&	\mathcal{C}[\,\cdot\,]	&	\mathcal{P}[\,\cdot\,]	&	\mathcal{T}[\,\cdot\,]	&	\mathcal{CPT}[\,\cdot\,]
		& \dual[\,\cdot\,]	&	\dual'[\,\cdot\,]	\\
	\hline\hline
		\hat\sigma_a	&	1	&	0	&	\hat\sigma_a^\dag	&	\hat\sigma_{-a}	&	\hat\sigma_a^\dag	&	\hat\sigma_{-a}
			&	\hat\mu_{a+\frac12}	&	\hat\mu^{\prime\dag}_{a-\frac12}	\\
		\hat\tau_a	&	0	&	0	&	\hat\tau_a^\dag	&	\hat\tau_{-a}	&	\hat\tau_a	&	\hat\tau_{-a}^\dag
			&	\hat\nu_{a+\frac12}	&	\hat\nu_{a-\frac12}^\dag	\\
		\hat\mu_b \;= \prod_{a<b}\hat\tau_a	&	0	&	1	&	\hat\mu_b^\dag	&	\hat\mu^{\prime\dag}_{-b}	&	\hat\mu_b	&	\hat\mu^{\prime}_{-b}
			&	\hat\sigma_{b+\frac12}	&	\mathcal{Q}^\textrm{dual}\hat\sigma_{b-\frac12}^\dag	\\
		\hat\mu'_b \;= \prod_{a>b}\hat\tau_a^\dag 	&	0	&	1	&	\hat\mu^{\prime\dag}_b	&	\hat\mu^\dag_{-b}	&	\hat\mu'_b	&	\hat\mu_{-b}
			&	\mathcal{Q}^\textrm{dual}\hat\sigma_{b+\frac12}	&	\hat\sigma_{b-\frac12}^\dag	\\
		\hat\nu_b = \hat\sigma_{b-\frac12}^\dag\hat\sigma_{b+\frac12}^\phd	&	0	&	0	&	\hat\nu_b^\dag	&	\hat\nu^\dag_{-b}	&	\hat\nu_b^\dag	&	\hat\nu^\dag_{-b}
			&	\hat\tau_{b+\frac12}	&	\hat\tau_{b-\frac12}^\dag	\\
		\mathcal{Q} \;= \prod_a\hat\tau_a	&	0	&	0	&	\mathcal{Q}^\dag	&	\mathcal{Q}	&	\mathcal{Q}	&	\mathcal{Q}^\dag
			&	\mathcal{Q}^\textrm{dual}	&	\mathcal{Q}^\textrm{dual\dag}	\\
		\mathcal{Q}^\textrm{dual} \;= \hat\sigma_{-\infty}\hat\sigma_{\infty}^\dag	&	0	&	0	&	\mathcal{Q}^\textrm{dual\dag}	&	\mathcal{Q}^\textrm{dual\dag}	&	\mathcal{Q}^\textrm{dual\dag}	&	\mathcal{Q}^\textrm{dual\dag}
			&	\mathcal{Q}	&	\mathcal{Q}^\dag	\\
	\hline
	%	\hat\beta_{L,2a-1}	&	2	&	1	&	\hat\beta_{L,2a-1}^\dag	&	\hat\beta_{R,-2a}	&	\omega\mathcal{Q}\hat\beta_{R,2a-1}^\dag	&
	%		&	\omega^2 \hat\sigma_a^\phd \hat\mu_a^\dag	&	\hat\beta_{L,2a-2}	\\
	%	\hat\beta_{L,2a}	&	2	&	1	&	\hat\beta_{L,2a}^\dag	&	\hat\beta_{R,-1-2a}	&	\omega\mathcal{Q}\hat\beta_{R,2a}^\dag	&
	%		&	\hat\sigma_{a+1}^\phd \hat\mu_a^\dag	&	\hat\beta_{L,2a-1}	\\
		\hat\beta_{L,c} \quad\textrm{[Eq.~\eqref{betaL}]}	&	2	&	1	&	\hat\beta_{L,c}^\dag	&	\hat\beta_{R,-1-c}	&	\omega\mathcal{Q}\hat\beta_{R,c}^\dag	&	\omega\mathcal{Q}^\dag\hat\beta_{L,-1-c}
			&	\omega\mathcal{Q}\hat\beta_{L,c+1}^\dag	&	\hat\beta_{L,c-1}	\\
	%	\hat\beta_{R,2a+1}	&	2	&	2	&	\hat\beta_{R,2a+1}^\dag	&	\hat\beta_{L,-2-2a}	&	\omega\mathcal{Q}\hat\beta_{L,2a+1}^\dag	&
	%		&	\hat\beta_{R,2a+2}	&	\hat\mu'_a \hat\sigma_a^\phd	\\
	%	\hat\beta_{R,2a}	&	2	&	2	&	\hat\beta_{R,2a}^\dag	&	\hat\beta_{L,-1-2a}	&	\omega\mathcal{Q}\hat\beta_{L,2a}^\dag	&
	%		&	\hat\beta_{R,2a+1}	&	\omega \hat\mu'_{a-1} \hat\sigma_a^\phd	\\
		\hat\beta_{R,c} \quad\textrm{[Eq.~\eqref{betaR}]}	&	2	&	2	&	\hat\beta_{R,c}^\dag	&	\hat\beta_{L,-1-c}	&	\omega\mathcal{Q}\hat\beta_{L,c}^\dag	&	\omega\mathcal{Q}^\dag\hat\beta_{R,-1-c}
			&	\hat\beta_{R,c+1}	&	\omega\mathcal{Q}\hat\beta_{R,c-1}^\dag	\\
	\end{array} \end{align*}
	\caption{%
		Symmetry properties of various operators in the three-state Potts model.  In the second and third columns $Q_3$ and $Q_3^{\rm dual}$ denote the $\Zth$ and $\Zth^{\rm dual}$ charges carried by each operator.  (Recall that the eigenvalues of the generators $\mathcal{Q}$ and $\mathcal{Q}^\textrm{dual}$ are given by $\omega^{Q_3}$ and $\omega^{Q_3^\textrm{dual}}$, respectively.)
	}
	\label{tab:OpSym}
\end{table}

\subsection{Lattice parafermions}
\label{sec:para}

The easiest and most powerful way of analyzing the Ising model is to rewrite the transfer matrix/quantum Hamiltonian in terms of Majorana fermion operators. One elegant way of defining these lattice fermions is by taking the product of adjacent order and disorder operators \cite{KadanoffCeva}.
The resulting fermionic Hamiltonian is of considerable interest in its own right (particularly when the fermions themselves comprise the physical degrees of freedom), providing a simple but profound example of a topologically nontrivial phase \cite{KitMajorana}.

{\em Parafermions} in the three-state Potts model naturally generalize the Majorana fermions in Ising.  These operators are obtained by performing a Fradkin-Kadanoff transformation \cite{FK,AlcarazKoberle} analogous to the Jordan-Wigner transformation of the Ising model. Although the resulting lattice parafermions are similarly formed by products of order and disorder operators, a crucial difference arises: As discussed in Sec.~\ref{sec:dual}, the Potts model allows for two distinct duality transformations and hence two types of disorder operators ($\hat\mu_a$ and $\hat\mu_a'$).
Consequently, one can define two classes of lattice parafermions as follows:\footnote{We employ slightly different conventions for the lattice parafermions here compared to Ref.~\cite{Mong}, to adhere more closely to conventions in the CFT literature.
The parafermions denoted by $\alpha_{R/L}$ in \cite{Mong} are related to those above via $\hat\alpha_{R,a} = \hat\beta_{R,a}^\dagger$, $\hat\alpha_{L,a} = \omega^2 \hat\beta_{L,a}^\dagger \mathcal{Q}$.}
%\begin{subequations}\begin{alignat}{3}
%	\hat\beta_{L,2a-1}  &\equiv  \omega^2 \hat\mu'_{a-1} \hat\sigma^\dag_a
%		&\quad&	= (\omega\hat\sigma_a^\dag\tau_a^\dag) \hat\tau_{a+1}^\dag \hat\tau_{a+2}^\dag \cdots \ ,
%\\	\hat\beta_{L,2a}  &\equiv  \hat\mu'_{a} \hat\sigma^\dag_a
%		&\quad&	= (\hat\sigma_a^\dag) \hat\tau_{a+1}^\dag \hat\tau_{a+2}^\dag \cdots \ ,
%\\	\hat\beta_{R,2a-1}  &\equiv  \hat\sigma^\dag_a \hat\mu^\dag_{a-1}
%		&\quad&	= \cdots \hat\tau_{a-2}^\dag \hat\tau_{a-1}^\dag (\hat\sigma_a^\dag) \ ,
%\\	\hat\beta_{R,2a}  &\equiv  \omega \hat\sigma^\dag_a \hat\mu^\dag_{a}
%		&\quad&	= \cdots \hat\tau_{a-2}^\dag \hat\tau_{a-1}^\dag (\omega^2\hat\tau_a^\dag\hat\sigma_a^\dag) \ .
%\end{alignat}\end{subequations}
\begin{subequations}\label{betaL}\begin{alignat}{3}
	\hat\beta_{L,2a-1}  &\equiv  \omega^2 \hat\mu'_{a-\frac12} \hat\sigma^\dag_a
		&\quad&	= (\omega\hat\sigma_a^\dag\tau_a^\dag) \hat\tau_{a+1}^\dag \hat\tau_{a+2}^\dag \cdots \ ,
\\	\hat\beta_{L,2a}  &\equiv  \hat\mu'_{a+\frac12} \hat\sigma^\dag_a
		&\quad&	= (\hat\sigma_a^\dag) \hat\tau_{a+1}^\dag \hat\tau_{a+2}^\dag \cdots \ ,
\end{alignat}\end{subequations}
and
\begin{subequations}\label{betaR}\begin{alignat}{3}
	\hat\beta_{R,2a-1}  &\equiv  \hat\sigma^\dag_a \hat\mu^\dag_{a-\frac12}
		&\quad&	= \cdots \hat\tau_{a-2}^\dag \hat\tau_{a-1}^\dag (\hat\sigma_a^\dag) \ ,
\\	\hat\beta_{R,2a}  &\equiv  \omega \hat\sigma^\dag_a \hat\mu^\dag_{a+\frac12}
		&\quad&	= \cdots \hat\tau_{a-2}^\dag \hat\tau_{a-1}^\dag (\omega^2\hat\tau_a^\dag\hat\sigma_a^\dag) \ .
\end{alignat}\end{subequations}
(To obtain the right-hand sides we employed the identity $\hat\sigma_a \hat\tau_a = \omega \hat\tau_a \hat\sigma_a$.)
By examining the expressions on the right we see that these classes differ in the orientation of the $\hat\tau_a^\dagger$ strings: the ``left'' parafermion operators $\hat\beta_L$ have a string going off to $+\infty$, while the ``right'' parafermions $\hat\beta_R$ have a string emanating to $-\infty$.
We call these operators \emph{semi-local} because they involve strings that are related to symmetry generators and hence commute with Hamiltonian terms far from the string termination.
These operators  are not independent, as the $\hat\beta_R$'s may be written in terms of the $\hat\beta_L$'s and ${\cal Q}$; nevertheless both representations are very useful to retain since they transform into one another under certain symmetries as discussed below.
In lattice parafermion language, the Hamiltonian~\eqref{eq:H} reads
\begin{align}\begin{split}
	H &= -J \sum_{\!b\text{ even}\!} \omega^2 \hat\beta_{L,b+1}^\dag \hat\beta_{L,b}^\phd - f \sum_{b\text{ odd}} \omega^2 \hat\beta_{L,b+1}^\dag \hat\beta_{L,b}^\phd + \hc
	\\	&= -J \sum_{\!b\text{ even}\!} \omega \hat\beta_{R,b+1}^\dag \hat\beta_{R,b}^\phd - f \sum_{b\text{ odd}} \omega \hat\beta_{R,b+1}^\dag \hat\beta_{R,b}^\phd + \hc .
	\label{Hparafermion}
\end{split}\end{align}
%For the remainder of this section we focus on the critical point with $J = f$.

Using properties of the original Potts-model operators and their duals, it is straightforward to derive the following relations,
\begin{subequations}\begin{align}
	\hat\beta_{L,b}^3 &= \hat\beta_{R,b}^3 = 1 ,
\\	\hat\beta_{L,b}\, \hat\beta_{L,c} &= \omega^{ \operatorname{sgn}(c-b)}\, \hat\beta_{L,c} \,\hat\beta_{L,b} ,
\\	\hat\beta_{R,b}\, \hat\beta_{R,c} &= \omega^{ \operatorname{sgn}(b-c)}\, \hat\beta_{R,c} \,\hat\beta_{R,b} ,
\\	\hat\beta_{L,b}\, \hat\beta_{R,c} &= \begin{cases} \omega^{(-1)^b} \hat\beta_{R,c} \,\hat\beta_{L,b} & b = c, \\ \hat\beta_{R,c}\, \hat\beta_{L,b} & b \neq c.\end{cases}
\end{align}\end{subequations}
Hence $\hat\beta_{R/L}$ comprise $\Zth$ generalizations of Majorana fermion operators.
It is particularly noteworthy that the $\hat\beta$'s do not anticommute off-site, but rather swapping their order acquires a phase factor of $\omega$ (or $\omega^2$).
%Thus these operators are naturally associated with fractional spin-$2/3$ excitations, further suggesting a relation to chiral parafermion fields at criticality.
One can also use the symmetry properties of $\hat\sigma_a$, $\hat\tau_a$ to back out the transformations of $\hat\beta_{R/L}$ under the two $\Zth$ symmetries, as well as $\mathcal{C}$, $\mathcal{P}$, and $\mathcal{T}$.
Table \ref{tab:OpSym} summarizes the results.
Notice that both parity and time-reversal swap the right and left parafermion representations, hinting that such operators are related to chiral fields in the continuum limit.

Understanding the properties of the parafermion operators under duality is essential to the following analysis.
Deducing these properties requires some care because, for example, even though the two factors in $\hat\beta_{R,2a-1} = \hat\mu_{a-\frac12}^\dag \hat\sigma^\dag_a$ act on separate sites and commute, their duals under $\dual$ overlap on site $a$ and fail to commute.
One therefore must define how duality acts on products of operators.
We require that
\begin{itemize}
	\setlength{\itemsep}{0ex}
	\item	the duality operator $\dual$ be linear;
	\item	$\dual[\hat{X}^\dag] = \dual[\hat{X}]^\dag$;
	\item	if $\hat{X}$ and $\dual[\hat{X}]$ are local operators, then duality is distributive over multiplication, i.e., $\dual[\hat{X}\hat{Y}] = \dual[\hat{X}] \dual[\hat{Y}]$;
	\item	$\dual^2$ amounts to translation by one Potts-model site.
\end{itemize}
Using these conditions, we find that\footnote{For instance, our requirements on duality imply the following,
\begin{align}\begin{split}
		\hat\sigma_{a+1}^\dag \hat\mu_a^\dag
		&=	\dual^2[ \hat\sigma_{a}^\dag \hat\mu_{a-1}^\dag ]
	=	\dual[ e^{i\theta} \hat\sigma_a^\dag \hat\mu_a^\dag ]
       = \dual[ e^{i\theta} \hat\sigma_a^\dag \hat\mu_{a-1}^\dag \hat\tau_{a}^\dag ]
	= e^{i\theta} \dual[ \hat\sigma_a^\dag \hat\mu_{a-1}^\dag ] \dual[ \hat\tau_{a}^\dag ]
	= e^{i\theta} (e^{i\theta} \hat\sigma_a^\dag \hat\mu_{a}^\dag) (\hat\sigma_{a+1}^\dag \hat\sigma_a^\phd) \\	
&= e^{2i\theta} \omega \hat\sigma_{a+1}^\dag \hat\mu_{a}^\dag\ .
\end{split}\end{align}
Hence $e^{2i\theta} \omega = 1$.  Setting $(\dual[\hat\beta_{R,2a-1}])^3 = 1$
yields $e^{i\theta} = \omega$, so that $\dual[\hat\beta_{R,2a-1}] = \hat\beta_{R,2a}\ .$}
\begin{subequations}\begin{align}
	\dual[\hat\beta_{R,2a-1}] &= \dual[ \hat\mu_{a-\frac12}^\dag \hat\sigma^\dag_a ] = \omega \hat\sigma_a^\dag \hat\mu_{a+\frac12}^\dag = \hat\beta_{R,2a} \ ,
\\	\dual[\hat\beta_{R,2a}] &= \dual[ \omega^2 \hat\mu_{a+\frac12}^\dag \hat\sigma^\dag_a ] = \hat\sigma_{a+1}^\dag \hat\mu_{a+\frac12}^\dag = \hat\beta_{R,2a+1} \ ,
\end{align}\end{subequations}
which may be summarized succinctly as\footnote{Applying duality twice translates the lattice parafermion operators by \emph{two} sites of the parafermion chain, i.e., $\dual^2[\hat\beta_{R,b}] = \hat\beta_{R,b+2}$, but this corresponds to translation by a single Potts site as required.}
$\dual[\hat\beta_{R,b}] = \hat\beta_{R,b+1}$.
Applying similar logic for the left parafermions yields
\begin{align}
	\dual'[\hat\beta_{L,b}] = \hat\beta_{L,b-1} \ .
\end{align}
Both transformations are obvious in hindsight, as swapping even and odd parafermion sites interchanges the $J$ and $f$ terms in the Hamiltonian~\eqref{Hparafermion}, precisely as duality should.
Determining the action of $\dual$ on left parafermion operators or $\dual'$ on right parafermion operators is more complicated, but the full results appear in Table~\ref{tab:OpSym}.
%For example, $\beta_{L,2a} = \hat\sigma_a^\dag \hat\mu'_{a}$ has a string going towards $+\infty$, whereas $\dual[\hat\beta_{L,2a}] = \hat\mu_a^\dag\hat\sigma_{a+1}$ has a string going off to $-\infty$.

%%%%%%%%%%%%%%%%%%%%%%%%%%%%%%%%%%%%%%%%%%%%%%%%%%%%%%%%%%%%%%%%%%%%%%%%%%%%%%%%
\section{The \texorpdfstring{$\Zth$}{Z3} parafermion conformal field theory}
\label{sec:CFT}

When the Hamiltonian~\eqref{eq:H} is tuned to the ferromagnetic critical point with $f = J$---which we assume hereafter---its long-distance physics is described by a well-studied conformal field theory \cite{BPZ} known as the three-state Potts or $\Zth$ parafermion CFT \cite{Dotsenko}.
For brevity we often call this the $\Zth$ CFT (although the theory exhibits a full $S_3$ permutation symmetry) \cite{FZ2}.
Conformal symmetry is infinite-dimensional in 1+1 dimensions, and so the resulting constraints allow many properties to be understood exactly.
In this section we review some properties of the $\Zth$ CFT.
We also describe how one can deduce the behavior of the many interesting fields under the Hamiltonian's discrete symmetries.

\subsection{Primaries}
\label{Primaries}

The $\Zth$ CFT has central charge $c = 4/5$ \cite{Dotsenko} and is a rational conformal field theory. The fundamental characteristic of a {rational} conformal field theory is that all the operators/states of the theory can be expressed in terms of a {finite} set of operators dubbed {primary fields}. That is, every state in the Hilbert state may be constructed by acting with a primary field and the generators of the (possibly extended) conformal algebra.

With appropriate boundary conditions, the left- and right-moving conformal symmetries are independent.
When space-time is written in terms of complex coordinates, the corresponding generators are the holomorphic and antiholomorphic parts of the energy-momentum tensor, respectively.
Thus one can decompose any field into representations of these independent symmetries.  A given field therefore can be characterized by left and right scaling dimensions ($h$, $\bar h$), so that its total scaling dimension is $h+\bar{h}$ while its conformal spin is $h-\bar{h}$.  Local fields possess integer conformal spin and exhibit correlators that remain invariant under $2\pi$ rotations; parafermions (and fermions for that matter) do not represent local fields in this sense.

The $\Zth$ CFT supports additional spin $\pm 3$ currents denoted $W$ and $\overline{W}$. It is then useful to extend the usual conformal (Virasoro) algebra by these generators to obtain what is known as the ``$\mathcal{W}_3$ algebra'' \cite{FZW3}.
This is the simplest non-trivial CFT with this symmetry algebra.
Fortunately, for our purposes here the intricacies of the extended algebra are largely unimportant.
All we need to know is the list of primary fields and that the field content can be generated by operator product expansions (OPEs) of the primaries with the left- and right-moving stress-energy tensors $T$, $\Tbar$, and with $W$, $\overline{W}$.
The ``descendant fields'' obtained in this fashion yield all the operators/states in the theory.

The chiral building blocks of the fields are known as primary chiral vertex operators \cite{Moorereview}; we call these chiral primaries for short.  All primary fields, both local and non-local, can be built from linear combinations of products of chiral and anti-chiral primaries. It is important to note that in any conformal field theory other than that of a free boson, this decomposition is non-trivial.  Some of the fields are not simply the product of holomorphic and antiholomorphic fields; they are the {\em sum} of such products.

The six \emph{local} primary fields of the $\Zth$ CFT have long been known \cite{Cardymod}.
A set of local fields has the property that all their correlators remain unchanged under $2\pi$ rotations of the system; i.e., their conformal spin $h-\bar{h}$ is an integer.
For a given CFT, there is not a unique such set. As with the Ising model \cite{BPZ}, in parafermion theories one can form a set of local fields containing either the spin \emph{or} the disorder field, but not both: the OPE of the two contains fractional powers of $z$ [cf.\ Eq.~\eqref{eq:smu_ope}].
By convention we view the spin field as local.  This choice uniquely determines the set of local primaries, which we denote by 1, $s$, $s^\dag$, $E$, $\psi\bar\psi$, and $\psi^\dag\bar\psi^\dag$.

There is of course the identity field, labeled $1$.
The spin fields $s(z,\zbar)$ and $s^\dag(z,\zbar)$ each have dimensions $(1/15,1/15)$, and correspond to the scaling limit of the spin operators $\hat\sigma_a$, $\hat\sigma_a^\dagger$ described above.
Charge conjugation $\mathcal{C}$ interchanges them, so they form a doublet under the $S_3$ symmetry~\cite{CDisorder}.
The energy field $E(z,\zbar)$ possesses dimensions $(2/5,2/5)$.  Perturbing the critical theory by this field describes the scaling limit of the three-state Potts model away from criticality with $f/J \neq 1$ \cite{Dotsenko}.
We denote the chiral primaries comprising $s(z,\zbar)$ as $\sigma$; a full labeling includes the fusion channels \cite{Moorereview}, but we will not need this information.
We likewise label the chiral primaries that are part of $s^\dagger$ and $E$ by $\sigma^\dag$ and $\epsilon$, respectively, with the antichiral primaries labeled as $\bar{\sigma}$, $\bar{\sigma}^\dagger$, and $\bar{\epsilon}$.

As opposed to the spin and energy fields, the remaining two primaries of conformal spin zero split into a simple product of holomorphic and antiholomorphic fields.
It is thus convenient to denote them in terms of this product as $\psi\bar{\psi}$ and $\psi^\dagger\bar{\psi}^\dagger$.
The chiral components are the holomorphic ``parafermion'' fields $\psi,\,\psi^\dagger$, each with dimensions $(2/3,0)$, and their antiholomorphic cousins $\bar{\psi}, \bar{\psi}^\dagger$ which have dimensions $(0,2/3)$ \cite{FZ2}.
While the parafermion fields are closely related to the scaling limit of the lattice parafermion operators described in the previous section, we will show later that this relationship is more subtle than one might naively anticipate.
Identifying the precise connection between such lattice and continuum operators is the central goal of this work.

We stress that $\sigma$, $\sigma^\dag$, and $\epsilon$ are not physical fields, in the sense they cannot be realized separately by local or semi-local lattice operators in the three-state quantum Potts chain.
For this reason expressions like $\sigma\bar\sigma$ are deceptive (though sometimes used in the literature).
Instead we write $s = \Phi_{\sigma\bar\sigma}$, $E = \Phi_{\epsilon\bar\epsilon}$, $s^\dag = \Phi_{\sigma\bar\sigma}^\dag = \Phi_{\sigma^\dag\bar\sigma^\dag}$, etc.
On the other hand, $\psi$ and $\psi^\dag$ are physical fields arising from the operator product expansion of the spin $s$ and disorder $\mu$ fields, each of which \emph{can} be realized on the lattice.
Moreover, by taking appropriately twisted boundary conditions in the parafermion conformal field theory, states corresponding to the chiral parafermion fields do occur \cite{Cardymod,GepnerQiu}.
This is why we can safely express the remaining local primaries as $\psi\bar\psi$, and $\psi^\dag\bar\psi^\dag$.

It is worth noting that there is still a finite number of primary fields here even if the symmetry algebra is not extended. These primary fields occur in the $c=4/5$ CFT with the ``diagonal'' modular invariant \cite{Cardymod}. This CFT describes the continuum limit of another lattice model sometimes known as the tetracritical Ising model, or the $A_5$ model in the nomenclature of Ref.~\cite{PasquierADE} (the three-state Potts model corresponds to $D_4$). The two CFT's are related by an orbifold \cite{DHVW}, which on the lattice amounts to a generalization of Kramers-Wannier duality \cite{FG}.
A nice illustration of the relation between these CFT's is given by constructing a field $\vv{X}{\epsilon}$ using the OPE of the energy field $E(z,\zbar)$ with the spin-3 current $W(z)$.
Precisely,
\begin{align}
	E(z,\zbar) W(0) &= \frac{E_1}{z^2} \vv{X}{\epsilon}(0,0) + \frac{E_2}{z} \partial_z \vv{X}{\epsilon}(0,0) + \dots ,
	\label{W_OPE}
\end{align}
where $E_1$ and $E_2$ are (known) constants \cite{FZW3}.
The field $\vv{X}{\epsilon}$ carries dimensions (7/5,2/5); the notation indicates that its chiral parts are comprised of chiral vertex operators in a sector labeled by $X$. Taking the OPE of $\overline{W}$ with $E$ yields another field $\vv{\epsilon}{X}$ with dimensions (2/5,7/5), while the OPE with both $W$ and $\overline{W}$ gives a field $\vv{X}{X}$ of dimensions (7/5,7/5). Since these fields cannot be constructed from $E$ with only the stress-energy tensor $T$, they are primary fields in the tetracritical Ising CFT.
In other words, $X$ is a descendant of $\epsilon$ when considering the entire $\mathcal{W}_3$ algebra but is not a descendant under the Virasoro algebra.
Indeed, it appears in the Kac table of the $c=4/5$ minimal conformal field theory~\cite{BPZ}.
Similarly, the $W$ field itself is a $\mathcal{W}_3$ descendant, but not a Virasoro descendant of the identity field---which is to say one can clearly construct $W$ from combinations of $1$ and $W$'s, but not with $T$'s alone.
Likewise, it also appears in the Kac table of primary fields.

\comment{As an aside, we note that the algebra of the $\Zth$ CFT may also be understood as an extension of the Virasoro algebra extended by parafermion fields $\psi$ and $\bar\psi$ \cite{FZ2}.
In general, $\Zth$ parafermions $\psi$ have scaling dimension $(2p/3 + M, 0)$ for integers $p>0$, $M\geq0$.
For the simplest case $p = 1$, $M = 0$ we have precisely the $\Zth$ CFT.
The $\Zth$ CFT at $c=4/5$ can be written as a coset theory $\mathrm{SU}(2)_3/\mathrm{U}(1)$; It appears that in the more general case the resulting CFTs can often be written as a coset theory of the form $G_k / \mathrm{U}(1)^{\operatorname{rank}(G)}$ for some group $G$ and level $k$ \cite{NoyvertParafermion}.%}
	\footnote{.}}

\begin{table}[t]
	\begin{align*} \begin{array}{c|cccccccc}
		$Field$	&	1	& W	&	\psi	&	\psi^\dag	&	\epsilon	&	X	&	\sigma	&	\sigma^\dag	\\
		\hline
		h	&	0	&	3	&	2/3	&	2/3	&	2/5	&	7/5	&	1/15	&	1/15	\\
%		\mathcal{C}[\,\cdot\,]	&	1	&	-W	&	\psi^\dag	&	\psi	&	\epsilon	&	-X	&	\sigma^\dag	&	\sigma \\
%		\mathcal{P}[\,\cdot\,]	&	1	&		&	e^{i\pi h}\bar{\psi}	&	e^{i\pi h}\bar{\psi}^\dagger	&	e^{i\pi h}\bar{\epsilon}	&		&	e^{i\pi h}\bar{\sigma}	&	e^{i\pi h}\bar{\sigma}^\dagger	\\
%		\mathcal{T}[\,\cdot\,]	&	1	&		&	\bar{\psi}^\dag	&	\bar{\psi}	&	\bar{\epsilon}	&		&	\bar{\sigma}^\dag	&	\bar{\sigma}
	\end{array} \end{align*}
	\caption{%
		The chiral (holomorphic) primaries under the Virasoro algebra, together with their scaling dimensions $h$.
		Note that under the extended $\mathcal{W}_3$ algebra the fields $W$ and $X$ are descendants of $1$ and $\epsilon$, respectively.
%		Also, $1 = \bar{1}$ if anyone asks.
%		\roger{The bottom two rows need fixing.}
	}
	\label{tab:CFTspin}
\end{table}

Table~\ref{tab:CFTspin} enumerates the set of holomorphic chiral primaries and their scaling dimensions.

\subsection{Parafermions}
\label{ParafermionsSection}

%\paul{We need to define ``physical'' -- the only definition I know if is to say that we know a lattice analogue. Thus perhaps we should move the next few paragraphs below?}
This subsection discusses the parafermion fields introduced above in greater depth.
First, however, it is worth briefly digressing on the nature of physical fields.
Not all physical fields need to be local; for example, the disorder field $\mu$ introduces a branch cut in space-time but proves to be very useful.
The complete set of physical operators in the CFT includes \emph{semi-local} fields that---like $\mu$---contain a string that is invisible to the stress-energy tensor $T$ far away.
%Following Sec.~\ref{sec:symmetries} we call fields semi-local if they, like $\mu$, contain a string that is invisible to the stress-energy tensor $T$ far away.
In operator language, such strings represent `half' of a symmetry generator, as exemplified by the lattice operators $\hat\mu_b$ and $\hat\mu'_b$ (which together yield $\hat\mu_b\hat\mu'_b = \mathcal{Q}^\dag$).
For identifying semi-local field combinations, it is useful to separate the holomorphic %$\mathcal{W}_3$
primary fields into two groups: $[1]$ contains $1$, $\psi$, and $\psi^\dag$, while $[\epsilon]$ contains $\epsilon$, $\sigma$, and $\sigma^\dag$.
Likewise, we divide the antiholomorphic fields into analogous groups denoted $[\bar 1]$ and $[\bar\epsilon]$.
The field $\Phi_{f\bar{f}}$ (along with its descendants) is then semi-local if $f \in [1]$ and $\bar{f} \in [\bar1]$, or if $f \in [\epsilon]$ and $\bar{f} \in [\bar\epsilon]$.
With the inclusion of semi-local fields the set of permissible primary fields expands beyond the six local primaries discussed in Sec.~\ref{Primaries}.
Thus, combinations such as $\Phi_{\psi} = \psi$, $\Phi_{\epsilon\bar\sigma}$, $\Phi_{\sigma\Xbar}$ are all acceptable, but it appears that neither $\Phi_{\psi\bar\epsilon}$ or $\Phi_{\epsilon\bar1}$ are physical as they involve fields from different sets.%
	\footnote{We note that if one builds a string from the charge-conjugation operator $\hat{C}$, there can be additional fields from the tetracritical Ising model, which are beyond the 6 chiral primaries of the $\Zth$ parafermion CFT \cite{CDisorder}.  They will not be considered in this paper.}

With this in mind we turn now to the parafermion operators, which are particularly important since, for example, they provide a simple way of understanding the appearance of topological properties \cite{para}. %A central question in this paper is the relation of the lattice definitions to the conformal fields.
To begin identifying the link between their lattice and continuum realizations, recall from Sec.~\ref{sec:para} that the lattice parafermion operators were defined as products of order and disorder operators, following Ref.~\cite{FK}.  Likewise, the holomorphic and antiholomorphic parafermion fields are naturally defined in conformal field theory by taking the operator product of order and disorder \emph{fields} \cite{FZ2},
\begin{subequations}\begin{align}
	s^\dag(z,\zbar) \mu(0,0) &\sim \frac{1}{(z\zbar)^{2/15}} z^{2/3}\psi(0) + {\rm other~terms} ,
\\	s^\dag(z,\zbar) \mu^\dag(0,0) &\sim \frac{1}{(z\zbar)^{2/15}}
		\zbar^{2/3}\bar\psi(0) + {\rm other~terms} .
\end{align}\end{subequations}
If only descendants of parafermion fields comprised the `other terms' above, then the identification of the lattice analogues of the parafermions would be obvious.  A main message of our paper is that this is not so.  Rather, both of these operator products contain another field coming from a different sector,
%\paul{Introduce $\vv{\sigma}{\epsilon}$, point out that it has dimension -1/3, differing from 2/3 by an integer.}
\begin{subequations}\label{eq:smu_ope}\begin{align}
	s^\dag(z,\zbar) \mu(0,0) &= \frac{1}{(z\zbar)^{2/15}}
		\big[ C_{1} z^{1/15}\zbar^{2/5} \Phi_{\sigma\bar\epsilon}(0,0) + C_2 z^{2/3}\psi(0,0) + \dots \big] ,
\\	s^\dag(z,\zbar) \mu^\dag(0,0) &= \frac{1}{(z\zbar)^{2/15}}
		\big[ C_1^\ast z^{2/5}\zbar^{1/15} \Phi_{\epsilon\bar\sigma}(0,0) + C_2^\ast \zbar^{2/3}\bar\psi(0,0) + \dots \big] ,
\end{align}\end{subequations}
where $C_1$, $C_2$ are constants and the ellipses denote subleading terms. %
%\footnote{
%	The moduli of the OPE coefficients are given by $|C_1|^2 = \frac{\Gamma(\frac{1}{5})\Gamma(\frac{3}{5})^3} {2\Gamma(\frac{4}{5})\Gamma(\frac{2}{5})^3}$ and $C_2^2 = \frac{1}{3}$.
% \roger{The second one I'm sure of, the first one is guess work.}}
The operators $\Phi_{\sigma\bar\epsilon}$ and $\Phi_{\epsilon\bar\sigma}$ are not discussed in Ref.~\cite{FZ2}, but there is no obvious reason why they should be absent.
For instance $\Phi_{\sigma\bar\epsilon}$ carries the same $\Zth$ and $\Zth^\textrm{dual}$ charge as the parafermion field $\psi$.
Moreover, they do indeed appear in the partition function with twisted boundary conditions (Eq.~(B3) in Ref.~\cite{GepnerQiu}).
Section \ref{sec:OPE} confirms the presence of these operators.
This is particularly important given that $\Phi_{\sigma\bar\epsilon}$ has a smaller scaling dimension than $\psi$ and thus constitutes the most singular term in the OPE's for the spin and disorder fields.
Consequently the identification of the lattice analogues of the parafermions is subtler than is might first appear and requires a careful analysis of the discrete symmetries.

\subsection{Symmetry properties of \texorpdfstring{$\Zth$}{Z3} CFT fields}
\label{sec:symm}

In Sec.~\ref{sec:symmetries} we reviewed the symmetries of the three-state quantum Potts chain, and the corresponding transformation properties of lattice operators in the theory (recall Table~\ref{tab:OpSym}).
Here we sketch how one can leverage those results to deduce the symmetry properties of CFT fields at criticality.
This exercise, the outcome of which appears in Table~\ref{tab:CFTSym}, will prove instrumental in allowing us to complete our infrared/ultraviolet correspondence below.
It is simplest to begin with the local spin and disorder fields $s$ and $\mu$, which by definition are the continuum limits of the spin and disorder operators $\hat\sigma_a$ and $\hat\mu_b$ (up to subleading corrections).
All symmetry properties of those fields can therefore be immediately read off from those of the lattice operators.
It turns out that this is the only link between the original Potts model and the CFT that we will need---the transformation properties of the remaining CFT fields can be inferred from consistency with OPE's.
Consider, for instance, Eqs.~\eqref{eq:smu_ope}.
%It is obvious from these OPE's that $\Phi_{\sigma\bar\epsilon}$ and $\psi$ carry the same (known) $\Zth$ and $\Zth^{\rm dual}$ charges as $s^\dagger \mu$, and with some care their transformations under $\mathcal{C}$, $\mathcal{P}$, $\mathcal{T}$, and the dualities follow as well.
It is obvious from these OPE's that $\psi$ carries the same (known) $\Zth$ and $\Zth^\textrm{dual}$ charges as $s^\dagger \mu$, and with some care their transformations under $\mathcal{C}$, $\mathcal{P}$, $\mathcal{T}$, and the dualities follow as well.
We also note that all the neutral fields ($Q_3 = Q_3^\textrm{dual} = 0$) arise from OPEs between $s$ and $s^\dag$ and thus their symmetry properties can be inferred from those of the spin fields.
The charge of $\Phi_{\sigma\bar\epsilon}$ also follows as it arises from the OPE of $E$ with the $\psi$ field.
In total this procedure allows one to fill in all rows of Table~\ref{tab:CFTSym}.

%Table~\ref{tab:CFTSym} gives a partial list of the physical fields in the critical theory, as well as their behavior under various symmetries.

\begin{table}[t]
	\begin{align*} \begin{array}{c|ccc|cc|ccccc}
		\textrm{Field} &	\textrm{primary}	&	\textrm{dim}	& \textrm{spin}	&	Q_3 & Q_3^\textrm{dual}
		&	\mathcal{C}[\,\cdot\,]	&	\mathcal{P}[\,\cdot\,]	&	\mathcal{T}[\,\cdot\,]	& \dual[\,\cdot\,]	&	\dual'[\,\cdot\,]	\\
	\hline\hline
		s = \Phi_{\sigma\bar\sigma}	&	\checkmark	&	2/15	&	0	&	1	&	0
			&	s^\dag	&	s	&	s^\dag	&	\mu	&	\mu^\dag	\\
		\partial_x s	&	&	17/15	&	\pm1	&	1	&	0
			&	\partial_x s^\dag	&	-\partial_x s	&	\partial_x s^\dag	&	\partial_x\mu	&	\partial_x\mu^\dag	\\
		\partial_t s	&	&	17/15	&	\pm1	&	1	&	0
			&	\partial_t s^\dag	&	\partial_t s	&	-\partial_t s^\dag	&	\partial_t\mu	&	\partial_t\mu^\dag	\\
		\psi\bar\psi	&	\checkmark	&	4/3		&	0	&	1	&	0
			&	\psi^\dag\bar\psi^\dag	&	\psi\bar\psi	&	\psi^\dag\bar\psi^\dag	&	\psi^\dag\bar\psi	&	\psi\bar\psi^\dag	\\
	\hline
		\mu = -\vv{\sigma^\dag}{\sigma}	&	\checkmark	&	2/15	&	0	&	0	&	1
			%&	\mu^\dagger	&	\mathcal{Q}\mu^\dag	&	\mu	&	(?)s	&	(?)s^\dag	\\
			&	\mu^\dagger	&	\mu^\dag	&	\mu	&	s	&	s^\dag	\\
	\hline
		\Phi_{\sigma\bar\epsilon}	&	\checkmark	&	7/15	&	-1/3	&	2	&	1
			&	\Phi_{\sigma^\dag\bar\epsilon}	&	-\Phi_{\epsilon\bar\sigma}	&	\Phi_{\epsilon\bar\sigma^\dag}	&	-\Phi_{\sigma^\dag\bar\epsilon}	&	-\Phi_{\sigma\bar\epsilon}	\\
		\psi	&	\checkmark	&	2/3	&	2/3	&	2	&	1
			%&	\psi^\dag	&	(\omega\mathcal{Q}^2)\bar\psi	&	(\omega\mathcal{Q})\bar\psi^\dag	&	(?)\psi^\dag	&	\psi	\\
			&	\psi^\dag	&	\bar\psi	&	\bar\psi^\dag	&	\psi^\dag	&	\psi	\\
	\hline
		\Phi_{\epsilon\bar\sigma}	&	\checkmark	&	7/15	&	1/3	&	2	&	2
			&	\Phi_{\epsilon\bar\sigma^\dag}	&	-\Phi_{\sigma\bar\epsilon}	&	\Phi_{\sigma^\dag\bar\epsilon}	&	-\Phi_{\epsilon\bar\sigma}	&	-\Phi_{\epsilon\bar\sigma^\dag}	\\
		\bar\psi	&	\checkmark	&	2/3	&	-2/3	&	2	&	2
			%&	\bar\psi^\dag	&	(\omega^2\mathcal{Q})\psi	&	&	\bar\psi	&	(?)\bar\psi^\dag	\\
			&	\bar\psi^\dag	&	\psi	&	\psi^\dag	&	\bar\psi	&	\bar\psi^\dag	\\
	\hline\hline
		1	&	\checkmark	&	0	&	0	&	0	&	0	&	1	&	1	&	1	&	1	&	1	\\
		E = \Phi_{\epsilon\bar\epsilon}	&	\checkmark	&	4/5	&	0	&	0	&	0
			&	E	&	E	&	E	&	-E	&	-E	\\
		\partial_x E	&	&	9/5	&	\pm1	&	0	&	0
			&	\partial_x E	&	-\partial_x E	&	\partial_x E	&	-\partial_x E	&	-\partial_x E	\\
		\partial_t E	&	&	9/5	&	\pm1	&	0	&	0
			&	\partial_t E	&	\partial_t E	&	-\partial_t E	&	-\partial_t E	&	-\partial_t E	\\
		\Phi_{X\bar\epsilon}	&	&	9/5	&	1	&	0	&	0
			&	-\Phi_{X\bar\epsilon}	&	\Phi_{\epsilon\Xbar}	&	-\Phi_{\epsilon\Xbar}	&	\Phi_{X\bar\epsilon}	&	-\Phi_{X\bar\epsilon}	\\
		\Phi_{\epsilon\Xbar}	&	&	9/5	&	-1	&	0	&	0
			&	-\Phi_{\epsilon\Xbar}	&	\Phi_{X\bar\epsilon}	&	-\Phi_{X\bar\epsilon}	&	-\Phi_{\epsilon\Xbar}	&	\Phi_{\epsilon\Xbar}	\\
		T	&	&	2	&	2	&	0	&	0	&	T	&	\Tbar	&	\Tbar	&	T	&	T	\\
		\Tbar	&	&	2	&	-2	&	0	&	0	&	\Tbar	&	T	&	T	&	\Tbar	&	\Tbar	\\
%			\vdots	&	&	14/5	&	0,\pm2	&	0	&	0	&	\vdots	&	\vdots	&	\vdots	&	\vdots	&	\vdots	\\[1.8mm]
		W	&	&	3	&	3	&	0	&	0	&	-W	&	\overline W	&	-\overline W	&	-W	&	W	\\
		\overline W	&	&	3	&	-3	&	0	&	0	&	-\overline W	&	W	&	-W	&	\overline W	&	-\overline W	\\
	\end{array} \end{align*}
	\caption{%
		Partial list of physical fields in the $\Zth$ CFT, group by $Q_3$ and $Q_3^\textrm{dual}$ charges.
		The second column indicates which fields are primary while the third and fourth list the scaling dimension $(h+\bar h)$ and spin $(h-\bar h)$ of each.  (A spin of $\pm 1$ means that the corresponding fields by themselves do not have well-defined spin; they can, however, be combined to yield fields with a spin of either $+1$ or $-1$, e.g., $\partial_z s$ and $\partial_{\zbar} s$)
		Remaining columns specify their symmetry properties.   (We note that factors of $\omega$, $\mathcal{Q}^\textrm{dual}$, and $\mathcal{Q}$ have been suppressed in the table.)
		%The vertical ellipses denote fields of dimensions $2 + 4/5$ which are not shown in the table.
	}
	\label{tab:CFTSym}
\end{table}

\section{Identifying local fields with lattice operators}
\label{sec:local}

We are now ready to begin addressing the most important issue of this paper---finding lattice operators in the critical three-state Potts chain that in the continuum limit yield particular fields in the corresponding $\Zth$ CFT.
Here we find lattice realizations of \emph{local} fields, which according to our previous definition are those that are realizable in terms of local lattice operators (without strings).
This case is therefore simpler than that of the non-local operators undertaken in the next section.
In what follows we construct lattice realizations of each local primary CFT field as well as the energy-momentum tensor and the dimension (2/5,7/5) and (7/5,2/5) operators $\Phi_{\epsilon \bar X}$ and $\Phi_{X \bar\epsilon }$.
We utilize the symmetries reviewed in Secs.~\ref{sec:symmetries} and~\ref{sec:CFT} (together with integrability in one case) as a guiding principle and present density-matrix-renormalization-group (DMRG) simulations that verify our results; for some details on the method see the Appendix.
%Section *** explores some implications of the findings developed here.

%As defined earlier, local fields are those that are realizable in terms of local lattice operators (without strings).  A consequence of this definition is that a set of local fields has the property that all of their correlators remain unchanged under $2\pi$ rotations of the system.  This in turn implies that operator products within the set contain coefficients that are functions only of $z^\alpha\zbar^\beta$, where the exponents $\alpha,\beta$ differ by integers.
%For a given CFT, there is not a unique such set. As with the Ising model \cite{BPZ}, in parafermion theories, one can form a set of local fields containing either the spin \emph{or} the disorder field, but not both: the OPE of the two contains fractional powers of $z$ [cf.\ Eq.~\eqref{eq:smu_ope}].
%By convention, we view the spin field as local.

Section~\ref{sec:CFT} discussed at length the local fields in the three-state Potts model that are both relevant and Lorentz invariant (rotationally invariant if the two dimensions are interpreted classically).
They consist of the identity 1, the spin field $s$ and its conjugate $s^\dag$, the energy field $E$, and the parafermion bilinears $\psi\bar\psi$ and $\psi^\dagger\bar\psi^\dagger$.
In CFT language, these form the primary fields of the extended symmetry algebra with conformal spin zero (i.e., the scaling dimensions for the constituent right- and left-moving components match).
As we will detail shortly, however, this list does not exhaust the local, relevant fields here; there are others with conformal spin 1.

\begin{figure}[tb]
	\centering
	\includegraphics[width=67mm]{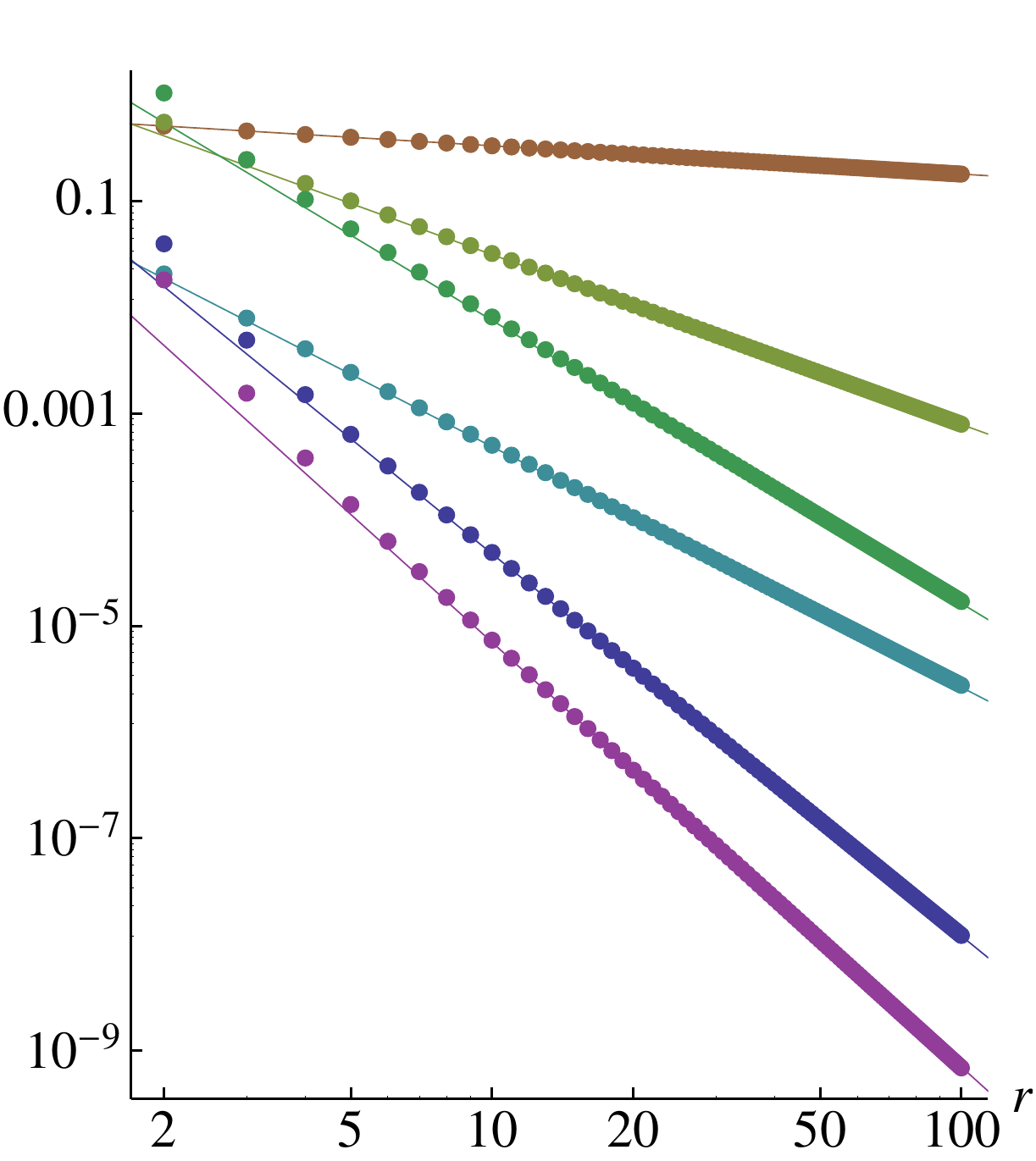}
	\;
	{\raisebox{40mm}{\begin{minipage}{83mm}
		\begin{align*}\renewcommand*{\arraystretch}{1.5}\begin{array}{c|cc}
			\hat{O}	&	 \renewcommand*{\arraystretch}{0.9}\begin{matrix}\textrm{Leading}\\\textrm{field}\end{matrix}	&	\textrm{dim }\Delta
		\\	\hline
			\hat\sigma_a	&	s(a)	&	2/15
		\\	\big( \hat\sigma_a^\phd \hat\sigma_{a+1}^\dag - \frac{\hat\tau_a^\phd + \hat\tau_{a+1}^\phd}{2} \big) + \hc	&	\Phi_{\epsilon\bar\epsilon}(a+\frac12)	&	4/5
		\\	\hat{\mathcal{B}}^\dag \quad\textrm{[Eq.~\eqref{eq:HmH0}]}	&	\psi\bar\psi(a+\frac12)	&	4/3
		\\	\omega\hat\sigma_a\hat\tau_a - \omega^2\hat\sigma_a\hat\tau_a^\dag	&	\partial_t s(a)	&	17/15
		\\	\frac{ \hat\tau_a^\phd + \hat\tau_{a+1}^\phd + 2\hat\sigma_a^\dag \hat\sigma_{a+1} }{ 2i } + \hc	&	\Phi_{X\bar\epsilon}(a+\frac12)	&	9/5
		\\	\hat S \pm \hat R \quad\textrm{[Eq.~\eqref{eq:latticeT}]}	&	T(a+\frac12)	&	2
		\end{array}\end{align*}
	\end{minipage}}}
	\caption{%
		Correlation function $\braket{\hat{O}(0) \hat{O}^\dag(r)}$ for various operators $\hat{O}$ listed in the table, displayed in the order in which they appear.
		All correlators exhibit power-law decay $\propto r^{-2\Delta}$ at large $r$.   The exponent $\Delta$ is given by the scaling dimension for the fields in the table's center column, confirming the lattice operator/field correspondences discussed in the text.
		%$\alpha = 4/15, 8/5, 8/3, 18/5, 4$.
		}
	\label{fig:CorrPlot_local}
\end{figure}

We already asserted that the spin field $s$ represents the continuum counterpart of the lattice operator $\hat\sigma_a$.
Nevertheless it is worth making some additional remarks regarding the inevitability of this identification.
Because $s$ is both the most relevant operator and breaks $\Zth$ symmetry, essentially any lattice operator with the latter property will have as its leading component the spin field in the continuum limit (except in special cases where other symmetries preclude this field from appearing).
As the nomenclature suggests, the simplest such operator is indeed $\hat\sigma_a$, which when diagonalized measures which of the three states is present on site $a$.
The identity $\hat\sigma_a^3=1$ is also consistent with the operator product expansion of $s$.
Finally, one can verify using DMRG that at criticality $\langle \hat\sigma_{a+r}^\dag \hat\sigma_a\rangle \sim r^{-4/15}$ for large $r$ as shown in Fig.~\ref{fig:CorrPlot_local}, in agreement with the expansion $\hat\sigma_a \sim s$.

Symmetry also allows one to identify the lattice analogue of the energy field $E$---so named because when added to the action it changes the temperature in the two-dimensional classical Potts model.
In the quantum chain, this simply corresponds to making $f\ne J$.
The energy field is neutral under both $\Zth$ and $\Zth^{\rm dual}$; even under $\mathcal{C}$, $\mathcal{T}$, and $\mathcal{P}$; but odd under both dualities ${\cal D}$ and ${\cal D'}$.
An appropriate combination of lattice operators sharing these symmetries is
\begin{align}
	\big( 2\hat\sigma_a^\phd \hat\sigma_{a+1}^\dag - \hat\tau_a^\phd - \hat\tau_{a+1}^\phd \big) + \hc \sim E \ .
	\label{eq:epep}
\end{align}
Note that adding such a term uniformly to the critical Hamiltonian indeed shifts the system off of criticality, and into a either the ferromagnetic or paramagnetic phase depending on the sign of the coupling constant.

The preceding identification of the spin and energy fields with lattice operators is fairly obvious and has long been known \cite{Dotsenko,AlcarazEnergy}.
The same is not true for the parafermion bilinears.
For reference, the analogous fermion bilinear $\psi\bar\psi$ in the Ising model is precisely the energy field---they are not independent perturbations unlike in the present context.
This is a succinct reason why the Ising model can be solved for any temperature; Hamiltonians composed of fermion bilinears are typically easily solvable, and hence their correlators can be readily computed.
In the three-state Potts and other parafermion models, a parafermion bilinear is a much more complicated object distinct from the energy field.
In fact, it does not even preserve the $\Zth$ symmetry (or more generally the $\mathbb{Z}_q$ symmetry in the $q$-state Potts case with $q>2$).
Moreover, it shares the the same symmetry properties as the spin field yet exhibits a larger scaling dimension.
When constructing a lattice analogue one therefore must effectively subtract off the bits that would scale onto the spin field.
Doing this by brute force seems prohibitively difficult.

Luckily, in all parafermion models integrability provides a means of finding a lattice analogue of the parafermion bilinears.
Consider a perturbed $\mathbb{Z}_q$ parafermion CFT described by the continuum Hamiltonian % Perturbing the parafermion CFT by these bilinears as
\begin{align}
	H = H_\textrm{CFT} -\lambda \int_x \psi(x) \bar\psi(x) + \hc
	\label{eq:Hpsipsi}
\end{align}
Here $H_\textrm{CFT}$ is the Hamiltonian at criticality while the $\lambda$ term breaks the $\mathbb{Z}_q$ symmetry (except in the Ising case).
This field theory is integrable and is analyzed in depth in Ref.~\cite{Fateev}.
Remarkably, there also exists an integrable $\mathbb{Z}_q$-breaking deformation \cite{KM} of the corresponding critical self-dual \emph{lattice} models \cite{FZ1}.
These lattice Hamiltonians are known explicitly, and the quantum Hamiltonians can be extracted by taking a particular anisotropic limit.
%[In Sec.~\ref{sec:Fibonacci} we also explain in depth how its continuum limit is indeed described by \eqref{eq:Hpsipsi} for $\lambda$ real and positive.]
For the three-state Potts model corresponding to $q = 3$, taking the limit very close to criticality yields
\begin{align}
	&	H = H^\textrm{crit}  -\Lambda\sum_a \left(\hat{\cal B}_a + \hat{\cal B}_a^\dagger\right)\ ;\\
	&	\hat{\cal B}_a^\dag \equiv \hat\sigma_a^\dag (2 - 3\omega^2\hat\tau_a - 3\omega\hat\tau_a^2) - 2(\hat\sigma_{a-1}\hat\sigma_a + \hat\sigma_a\hat\sigma_{a+1}) ,
	\label{eq:HmH0}
\end{align}
with $H^\textrm{crit}$ corresponding to the critical Potts chain and $\Lambda$ a `small' parameter.

Comparing with Eq.~\eqref{eq:Hpsipsi}, it is natural to identify the lattice analogue of $\psi(x) \bar\psi(x) + \hc$ as $\hat{\cal B}+\hat{\cal B}^\dag$.
The part corresponding to $\psi(x) \bar\psi(x)$ alone (rather than to its Hermitian conjugate) is fixed uniquely by the $\Zth$ charge, giving
\begin{align}
	\hat{\cal B}_a \;\sim\; \psi\bar\psi .
	\label{Bexpansion}
\end{align}
The combination of lattice operators on the left indeed satisfies the same symmetry properties as $\psi\bar\psi$ given in Table~\ref{tab:CFTSym}---but importantly also those of the spin field (which again has a smaller scaling dimension).
To solidify this correspondence it is thus crucial to verify that correlations of $\hat{\cal B}_a$ obey the proper scaling relation.
We have checked using DMRG on the Potts chain that the two-point function $\braket{\hat{\cal B}^\dagger_{a+r} \hat{\cal B}_a}$ evaluated in the ground state falls off as $r^{-8/3}$; see Fig.~\ref{fig:CorrPlot_local}.
Thus the lattice operator $\hat{\cal B}_a$ indeed has dimension $4/3$---just like $\psi\bar\psi$---which completes the proof of Eq.~\eqref{Bexpansion}.
%\paul{This plot should contain both the 8/3 part and the 18/5 part}.

We have now found lattice analogues of all the relevant rotationally invariant fields in the three-state Potts model. There are, however, other interesting fields that are chiral---i.e., {not} parity-invariant, or rotationally invariant in the two-dimensional classical model---yet possess integer conformal spin and hence remain local.
As described in Sec.~\ref{sec:CFT} there is an integer-spin symmetry generator $W$, and taking the operator product of this with rotationally invariant fields can yield chiral fields with integer spin.

The most interesting of these are the chiral self-dual fields $\Phi_{X\bar\epsilon}$ and $\Phi_{\epsilon \Xbar}$, which come from the leading term in the OPE of $W$ with the energy field [see Eq.~\eqref{W_OPE}].
These fields respectively possess left and right dimensions of $(2/5,7/5)$ and $(7/5,2/5)$, and each remains invariant under one of the dualities $\dual,\dual'$ but is odd under the other.%
	\footnote{Since $\dual$ takes $\hat\beta_L$ to $\hat\beta_L^\dag$ (with a phase), it must take $\psi$ to $\psi^\dag$ while leaving the antiholomorphic sectors unchanged.
		In effect, $\dual$ swaps $\psi$ and $\psi^\dag$ (and also $\sigma$ and $\sigma^\dag$) and thus behaves as charge conjugation acting only in the holomorphic sector.
		For consistency, charge conjugation must also be accompanied by the negation of the $\mathcal{W}_3$ descendant fields $W$ and $X$.
		We can therefore distinguish these fields from their parent fields $1$, $\epsilon$ with appropriate analysis under both dualities, which are listed in Table~\ref{tab:CFTSym}.}
Constructing the simplest lattice operators with these properties suggests the following correspondence \cite{CardyPotts}
\begin{subequations} \label{eq:eXlattice} \begin{align}
	\frac{ \hat\tau_a^\phd + \hat\tau_{a+1}^\phd + 2\hat\sigma_a^\dag \hat\sigma_{a+1} }{ 2i } + \hc  \;&\sim\;  \Phi_{X\bar\epsilon} \ ,
\\	\frac{ \hat\tau_a^\phd + \hat\tau_{a+1}^\phd - 2\hat\sigma_a^\dag \hat\sigma_{a+1} }{ 2i } + \hc  \;&\sim\;  \Phi_{\epsilon\Xbar} \ .
\end{align} \end{subequations}
To confirm this identification, we used DMRG to compute the two-point function of the lattice operators above in the ground state of the Potts chain.
Our data plotted in Fig.~\ref{fig:CorrPlot_local} show that indeed the operators in Eqs.~\eqref{eq:eXlattice} have dimension $9/5$---consistent with the field expansions on the right-hand side.

An additional convincing argument for these lattice analogues follows from the two-dimensional classical chiral Potts lattice model \cite{CardyPotts}---a chiral-symmetry-breaking deformation of the usual Potts model that preserves self-duality, is integrable, and has striking non-renormalization properties~\cite{Perkoverview}.
By analyzing its behavior under 90-degree rotations of the square lattice, it follows that the perturbing operator has conformal spin $\pm 1$ in the continuum limit, as do $\Phi_{X\bar{\epsilon}}$ and $\Phi_{\epsilon \Xbar}$ \cite{CardyPotts}.
In the quantum Hamiltonian limit this deformation arises upon including phase factors in the ferromagnetic and transverse-field couplings, e.g., $\hat\tau_a^\dag + \hat\tau_a \rightarrow e^{i\phi} \hat\tau_a^\dag + e^{-i\phi}\hat\tau_a$ in the latter.
Adding either of the lattice operators in Eqs.~\eqref{eq:eXlattice} uniformly to the Hamiltonian introduces precisely such chiral phases, which makes the field correspondence above rather natural.
% Just to make sure, D takes nu = sd(a) s(a+1) to tau, D also takes Xe to itself, and eX to -eX.

Finally, using the combination of parity $\mathcal{P}$, time-reversal $\mathcal{T}$, and charge conjugation $\mathcal{C}$, we deduce lattice operators corresponding to the left- and right-moving pieces of the stress-energy tensor, $T$ and $\Tbar$. Since the continuum Hamiltonian is proportional to $\int dx (T+\Tbar)$, one expects that
\begin{align}
	\hat{S}_{a+1/2} \equiv (-\sigma_a\sigma_{a+1}^\dag - \frac{\tau_a + \tau_{a+1}}{2} + \hc) - E_0 \;\sim\; T + \Tbar,
	\label{eq:TplusTbar}
\end{align}
where $E_0$ is the ground state energy per site at criticality; $E_0/J = -\frac{4}{3} - \frac{2\sqrt{3}}{\pi} \approx -2.4360$ \cite{YangYang,AlcarazEnergy}.
Subtracting the ground-state energy cancels the identity piece from the field expansion on the right side of Eq.~\eqref{eq:TplusTbar}. This identification is consistent with all discrete symmetries. Finding $T-\Tbar$ is trickier.
Notice first that there are eight fields in Table \ref{tab:CFTSym} that carry no charges ($Q_3 = Q_3^\textrm{dual} = 0$) and possess scaling dimensions $\leq 2$.
However, only the combination $T-\Tbar$ is odd under both parity and time-reversal but invariant under charge conjugation.
The following lattice operator also satisfies such symmetry properties and thus should include $T - \Tbar$ as its most relevant term,
\begin{align}
	\hat{R}_{a+1/2} \equiv
	2\big( \hat\sigma_a\hat\tau_a \hat\sigma_{a+1}^\dag\hat\tau_{a+1} - \hat\sigma_a^\dag\hat\tau_a \hat\sigma_{a+1}\hat\tau_{a+1} \big) + \hc
		\;\sim\; T-\Tbar. %T(x+1/2) - \Tbar(x+1/2)
	\label{eq:TminusTbar}
\end{align}
%Combined with the stress-energy tensor $\hat{S}_{x+1/2} = (-\sigma_x\sigma_{x+1}^\dag - \frac{\tau_x + \tau_{x+1}}{2} + \hc) - E_0 \sim T + \Tbar$, where $E_0$ is the ground state energy per site (assuming $J = f = 1$\footnote{$E_0 \approx -2.436$.}), we can get $T$ and $\Tbar$ individually.
As an additional check, we found using the DMRG that the cross correlator $\braket{\hat{R}_b\hat{S}_{b+r}}$ vanishes, consistent with the field assignments in Eqs.~\eqref{eq:TplusTbar} and \eqref{eq:TminusTbar}.

The chiral stress-energy tensor therefore appears through {some} linear combination of $\hat{R}$ and $\hat{S}$.
Remarkably, the DMRG numerics show that $\braket{\hat{R}_b\hat{R}_{b+r}} = \braket{\hat{S}_b\hat{S}_{b+r}}$ within numerical error;
hence the constant of proportionality for Eq.~\eqref{eq:TminusTbar} is equal in magnitude to that of Eq.~\eqref{eq:TplusTbar}.
Combining the two lattice operators above then allows us to extract $T$ and $\Tbar$ individually, i.e.,
\begin{align}
	\hat{S} \pm \hat{R} \;\sim\; T \textrm{ or } \bar{T} .
	\label{eq:latticeT}
\end{align}
We cannot determine which sign corresponds to $T$, $\bar{T}$ from two-point correlators alone, but it is in principle possible to distinguish them via three-point correlators, similar to the technique explained in Sec.~\ref{sec:OPE}.
We also note that $\sum_b\hat{R}_b$ defined above is not invariant under duality.
Thus we may define an alternative operator $\hat{R}_a' = \dual[\hat{R}_{a-1/2}]$ whose leading field is also $T - \bar{T}$.
This illustrates the fact that the lattice analogue of a given field is not unique; in fact, we have found yet another lattice analogue of $T-\bar{T}$:
\begin{align}
	\omega\hat\sigma_a^\dag\hat\tau_a^\dag\hat\sigma_{a+1} + \omega^2\hat\sigma_a\sigma_{a+1}^\dag\hat\tau_{a+1} - \omega\hat\sigma_a\hat\sigma_{a+1}^\dag\hat\tau_{a+1}^\dag - \omega^2\hat\sigma_a^\dag\hat\tau_a\sigma_{a+1} + \hc  \;\sim\;  T - \bar{T} .
\end{align}

To get the lattice operator dominated by a spatial derivative of any field, we can simply take the lattice difference of the corresponding operator.
For example, $\hat\sigma_{a+1} - \hat\sigma_a \sim \partial_x s$.
Taking the commutator with the Hamiltonian Eq.~\eqref{eq:H} similarly yields the time derivative.
Thus, $[H, \hat\sigma_a] \propto \omega\hat\sigma_a\hat\tau_a - \omega^2\hat\sigma_a\tau_a^\dag \sim \partial_t s$, as displayed and confirmed in Fig.~\ref{fig:CorrPlot_local}.

\section{Chiral CFT fields and semi-local lattice operators}
\label{sec:chiral}

In this section we turn to the more difficult task of identifying lattice analogues of \emph{non-local} CFT fields.  The chiral parafermions $\psi, \bar\psi$ comprise the most interesting examples, since the corresponding continuum fields are holomorphic and anti-holomorphic, of conformal spin $\pm 2/3$. We explain how to ``separate'' them from the non-holomorphic fields $\Phi_{\sigma\bar\epsilon}$ and $\Phi_{\epsilon\bar\sigma}$ of conformal spin $\mp 1/3$.  Combined with the correspondences derived in the previous section, we will then have obtained lattice analogues of \emph{all} physical fields in the $\Zth$ CFT with scaling dimension $\leq 2$.

As described in the introduction, the non-local field case is interesting both for formal and for physical reasons.
The solution is nontrivial since the expansion of any local combination of the original $\hat\sigma_a$ and $\hat\tau_a$ Potts operators can be treated using methods of the previous section, and is guaranteed to involve purely local fields.
Evidently these are the `wrong' degrees of freedom to utilize when seeking realizations of non-local fields on the lattice.
Equation~\eqref{eq:smu_ope} provides a clue for how to remedy this problem.
Namely, since the parafermion fields are defined through the OPE of spin and disorder fields, the `correct' representation should involve products of the original Potts operators and their duals (like the lattice parafermion operators introduced in Sec.~\ref{sec:para}).
These combinations are non-local---though they are semi-local---and hence need not involve local CFT fields in their expansion.

Let us then deduce the field expansion of the lattice parafermion $\hat\beta_{R,a}$ defined in Eqs.~\eqref{betaR}.
Comparing the $\Zth$ and $\Zth^{\rm dual}$ charges listed in Tables~\ref{tab:OpSym} and~\ref{tab:CFTSym}, we know that the only physical fields that can appear in the expansion are $\bar\psi$ (scaling dimension $2/3$), $\Phi_{\epsilon\bar\sigma}$ (dimension $7/15$), and their descendants.
Since $\Phi_{\epsilon\bar\sigma}$ has the smallest scaling dimension in this set, finding an appropriate lattice operator that yields $\bar\psi$ as its leading piece naively seems difficult.
A very similar state of affairs arose in Sec.~\ref{sec:local} when we determined the lattice analogue of the (local) parafermion bilinear $\psi\bar\psi$; recall Eq.~\eqref{Bexpansion}.
In that context, $\psi \bar\psi$ also carried the same charges as another, more relevant field: $s$.
Finding a lattice realization of the former then required using integrability to effectively eliminate a contribution from the latter.
Quite surprisingly, the situation is much cleaner for the single chiral parafermion $\bar\psi$, as symmetry alone allows one to isolate $\psi$ and $\Phi_{\epsilon\bar\sigma}$.
%Given its $Q_3$ and $Q_3^\textrm{dual}$ charges (see Tab.~\ref{tab:CFTSym}), we know it must contain components of $\psi$ and $\sigma\bar\epsilon$, with scaling dimensions $2/3$ and $7/15$ respectively (and their descendants).
%However, since the latter is the more relevant field of the two, it will generically be present.

Duality symmetry provides an appealing way to see this.
%Recall that the $\dual'$ transformation takes the field $\epsilon\bar\epsilon$ to $-\epsilon\bar\epsilon$, whereas $1\bar1$ remains invariant.
The crucial point is that $\bar\psi$ is even under $\dual$ whereas $\Phi_{\epsilon\bar\sigma}$ is odd.
A simple way of understanding why comes from the fusion rule $\bar\psi \times \bar\epsilon \sim \bar\sigma$.
The quantum number for $\Phi_{\epsilon\bar\sigma}$ therefore must be the composite of those for $\bar\psi$ and the energy field $E = \Phi_{\epsilon\bar\epsilon}$.
The latter is odd under $\dual$, because adding it to the action perturbs the system into the high- or low-temperature phases, with duality exchanging the two. Thus $\Phi_{\epsilon\bar\sigma}$ and  $\bar\psi$
indeed have opposite signs under duality.

\begin{figure}[tb]
	\centering
	\includegraphics[width=0.5\textwidth]{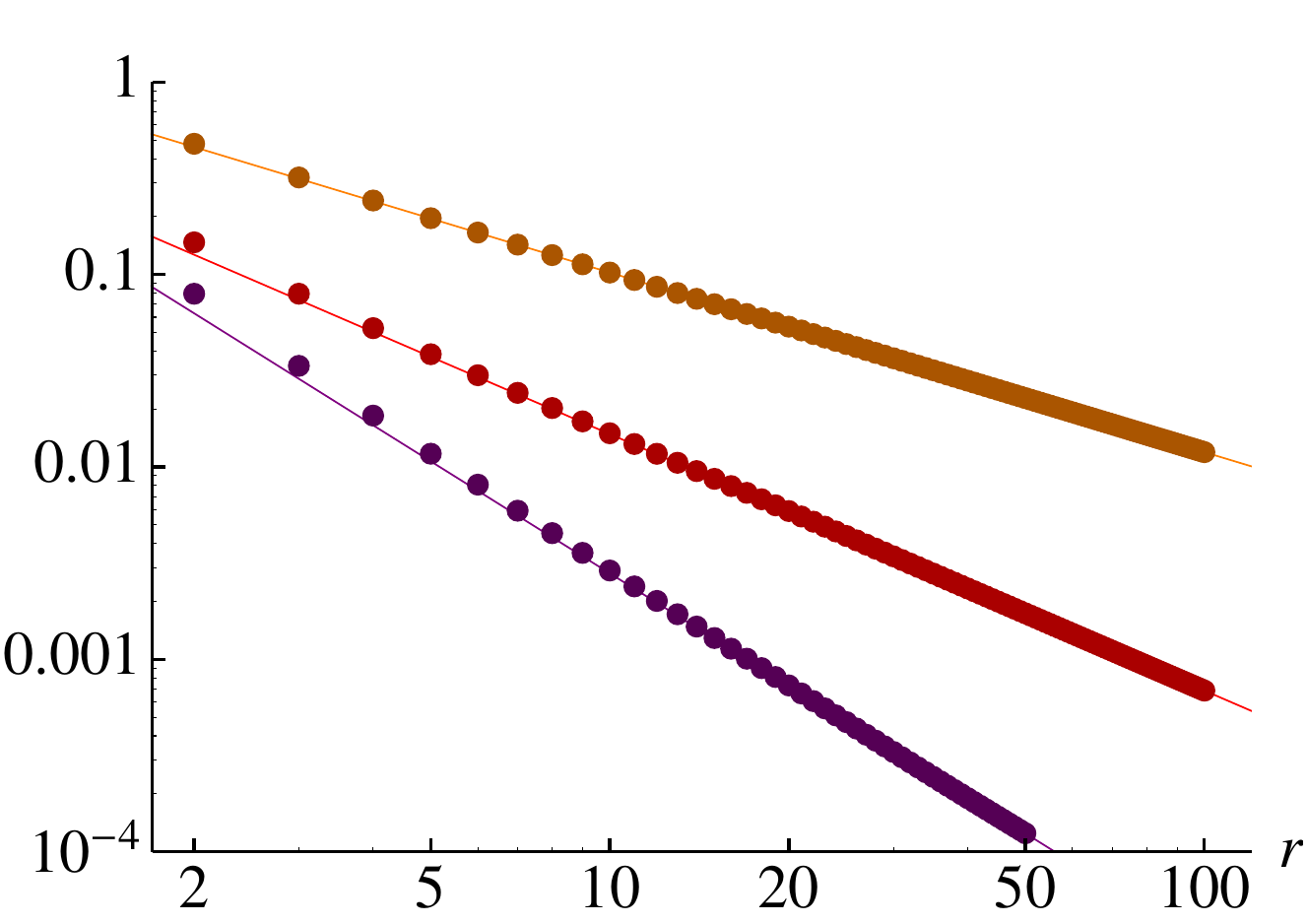}
	\caption{%
		Log-log plot of two-point correlation functions involving lattice operators $\hat A_{a}\equiv \hat\beta_{R,2a-1}+ \hat\beta_{R,2a}$ and $\hat B_{a}\equiv \hat\beta_{R,2a-1}- \hat\beta_{R,2a}$ from Eqs.~\eqref{eq:OpPsi} and \eqref{eq:OpSigmaE}.  Red data points correspond to $\langle \hat A_a \hat A_{a+r}^{\dag}\rangle$ which exhibits algebraic decay $\propto r^{-4/3}$, confirming the field correspondence $\hat A_a \sim \bar\psi$ derived in the main text.  Similarly, orange data represent $\langle \hat B_a \hat B_{a+r}^{\dag}\rangle$; this correlator decays as $r^{-14/15}$ consistent with the identification $\hat B_a \sim \Phi_{\epsilon\bar\sigma}$.   The cross correlator $\langle \hat A_a \hat B_{a+r}^{\dag}\rangle$ decays as $r^{-(1+14/15)}$ as shown by the purple points. 	}
	\label{fig:CorrPlot_nonlocal}
\end{figure}

On the lattice, $\dual$ takes the sum $S_\textrm{odd} = \sum_{a\in\mathbb{Z}} \hat\beta_{R,2a-1}$ to $S_\textrm{even} = \sum_{a\in\mathbb{Z}} \hat\beta_{R,2a}$ and vice-versa, so that the superpositions $S_{\rm odd} \pm S_{\rm even}$ possess eigenvalue $\pm1$ under $\dual$.  Thus it is natural to anticipate the expansions $\sum_{a\in\mathbb{Z}}(\hat\beta_{R,2a-1}+\hat\beta_{R,2a}) \sim \int_x\! \bar\psi$ and $\sum_{a\in\mathbb{Z}}(\hat\beta_{R,2a-1}-\hat\beta_{R,2a}) \sim \int_x\! \Phi_{\epsilon\bar\sigma}$, which in turn suggest the following lattice-operator/field correspondences:
\begin{alignat}{3}
	\hat\beta_{R,2a-1} + \hat\beta_{R,2a}
		\;&=\; \cdots \hat\tau_{a-2}^\dag \hat\tau_{a-1}^\dag (\hat\sigma_a^\dag + \omega^2\hat\tau_a^\dag\hat\sigma_a^\dag)
		\quad&&\sim\;  \bar\psi \ ,
	\label{eq:OpPsi}\\
	\hat\beta_{R,2a-1} - \hat\beta_{R,2a}
		\;&=\; \cdots \hat\tau_{a-2}^\dag \hat\tau_{a-1}^\dag (\hat\sigma_a^\dag - \omega^2\hat\tau_a^\dag\hat\sigma_a^\dag)
		\quad&&\sim\;  \Phi_{\epsilon\bar\sigma} \ .
	\label{eq:OpSigmaE}
\end{alignat}
One can recast these equations more compactly as
\begin{equation}
	\hat\beta_{R,a} \sim c_1\bar\psi + c_2 (-1)^a \Phi_{\epsilon\bar\sigma} ,
	\label{eq:betaRexpansion}
\end{equation}
where $c_1$ and $c_2$ denote non-universal constants.
An essentially identical argument utilizing $\dual'$ instead of $\dual$ suggests an analogous expansion for $\hat\beta_{L,a}$:
\begin{align}
	%\omega \mathcal{Q}^2 \hat\beta_{L,a} \sim c_1\psi + c_2 (-1)^a \Phi_{\sigma\bar\epsilon} .
	\hat\beta_{L,a} \sim d_1\psi + d_2 (-1)^a \Phi_{\sigma\bar\epsilon} .
	\label{eq:betaLexpansion}
\end{align}
%Since the CFT field opeators have string going towards $-\infty$ (by our convention) and $\hat\beta_L$'s have strings going to $+\infty$, the factor $\omega\mathcal{Q}^2$ is merely bookkeeping to account for this discrepency.
The remaining symmetries $\mathcal{C}$, $\mathcal{P}$, and $\mathcal{T}$ simply enforce that the coefficients $c_j$, $d_j$ are real and satisfy $c_j = d_j$.
%\jason{This last statement is worth an independent check.}\roger{good luck.}
Figure~\ref{fig:CorrPlot_nonlocal} plots the two-point correlation function of the lattice operator in Eq.~\eqref{eq:OpPsi} as a function of separation $r$ obtained from DMRG.
The correlator exhibits algebraic decay proportional to $r^{-4/3}$ in precise agreement with the right-hand side (again, $\bar\psi$ has scaling dimension $2/3$).

We have thus succeeded in identifying a lattice analogue of the chiral parafermion fields $\psi$ and $\bar\psi$.  To further confirm that we have ``separated'' $\psi$ from $\Phi_{\epsilon\bar\sigma}$, we also plot in Fig.~\ref{fig:CorrPlot_nonlocal} the two-point correlator for the lattice operator in Eq.~\eqref{eq:OpSigmaE}.  This indeed decays as $r^{-14/15}$, consistent with the $7/15$ scaling dimension of $\Phi_{\epsilon\bar\sigma}$.

We close this section with some comments.

(\textit{i}) At the critical point $J = f$ the lattice parafermion Hamiltonian~\eqref{Hparafermion} takes the form of a uniform chain with larger translational symmetry than the original Potts model.
That is, the unit cell for the parafermion chain is effectively halved at criticality---a consequence of duality symmetry.
From this point of view Eqs.~\eqref{eq:betaRexpansion} and \eqref{eq:betaLexpansion} indicate that the dominant piece of the lattice parafermion operators occurs at momentum $\pi$, with a subleading piece at zero momentum.
When averaged over many lattice spacings the dominant momentum-$\pi$ component gets `smeared out', leaving only the chiral parafermions.

(\textit{ii}) Additional useful insight into the expansions can be obtained from $\mathcal{CPT}$ symmetry.
Notice from Table~\ref{tab:OpSym} that the combination of $\mathcal{CPT}$ takes `odd' parafermions $\hat\beta_{R,2\mathbb{Z}+1}$ to `even' parafermions $\hat\beta_{R,2\mathbb{Z}}$, modulo the uniform phase factor $\omega\mathcal{Q}^\dag$.
Hence the left sides of Eqs.~\eqref{eq:OpPsi} and~\eqref{eq:OpSigmaE} are both eigenvectors under the action of $\mathcal{CPT}$, with eigenvalues that differ by a minus sign.
%	\footnote{Technically only true at site 0, since $\mathcal{P}$ takes site $a$ to $-a$.}
Acting on fields, $\mathcal{CPT}$ behaves similarly to a $\pi$ rotation in the space-time plane.
Since the spins of $\bar\psi$ and $\Phi_{\epsilon\bar\sigma}$, $-2/3$ and $1/3$ respectively, differ by 1, the right sides of Eqs.~\eqref{eq:OpPsi} and~\eqref{eq:OpSigmaE} also incur a relative minus sign under $\mathcal{CPT}$.
%In short, the momentum for the $\sigma\bar\epsilon$ component in the expansion can equivalently be understood either as a consequence of duality \emph{or} spin.

(\textit{iii}) As in all of our lattice/field correspondences, Eqs.~\eqref{eq:OpPsi} and \eqref{eq:OpSigmaE} hold up to subleading terms.
Both lattice operators contain further contributions consisting of symmetry-allowed descendants of $\bar\psi$ and $\Phi_{\epsilon\bar\sigma}$.
Interestingly, $\mathcal{CPT}$ symmetry allows us to also extract the second-most-relevant contribution in each field expansion.
The subleading term in Eq.~\eqref{eq:OpPsi} contains first descendants of $\Phi_{\epsilon\bar\sigma}$, such as $\partial_x\Phi_{\epsilon\bar\sigma}$, $\partial_t\Phi_{\epsilon\bar\sigma}$, and $\Phi_{X\bar\sigma}$, with scaling dimensions $22/15$, while the subleading terms in Eq.~\eqref{eq:OpSigmaE} are $\partial_x\bar\psi$ and $\partial_t\bar\psi$, both with scaling dimensions $5/3$.
Therefore, the cross-correlator of the two expression would be dominated by $\partial\Phi_{\epsilon\bar\sigma}$ from Eq.~\eqref{eq:OpPsi} and $\Phi_{\epsilon\bar\sigma}$ from Eq.~\eqref{eq:OpSigmaE}, decaying as $r^{-29/15}$.
Figure~\ref{fig:CorrPlot_nonlocal} shows agreement with this assessment.
%\jason{Is there a typo here?  $\Phi_{X\bar\sigma}$ can't appear as a subleading term in both cases since the left-hand sides have different symmetry properties...}

(\textit{iv}) We should emphasize that it is generally \emph{not} possible to infer the lattice analogue of chiral fields (e.g., $\psi$) from simply knowing the lattice analogue of local fields composed of left- and right-movers (e.g., $\psi \bar\psi$).
This should be clear by contrasting Eqs.~\eqref{Bexpansion} and \eqref{eq:OpPsi}.
Even simpler examples arise from products of left and right lattice parafermion operators; for instance, up to conserved quantities we have $\omega \hat\beta_{R,2a}\hat\beta_{L,2a} \propto \hat\sigma_a \sim s(a)$.
A naive reading of these relations would incorrectly suggest that $\hat\beta_{R,2a} \sim \bar\sigma$ and $\hat\beta_{L,2a} \sim \sigma$, where $\sigma$, $\bar\sigma$ are the chiral components of the spin field $s = \Phi_{\sigma\bar\sigma}$.\footnote{Recently, Refs.~\cite{VaeziKim,VaeziFib} also explored the correspondence between `high-energy' operators and CFT fields in parafermion theories.
		Differences between their results, and those obtained here and in Ref.~\cite{Mong}, can partly be attributed to the application of such logic.
		Moreover, we stress the importance of considering all discrete symmetries of the critical theory when identifying ultraviolet analogues of critical fields.  Failure to do so can lead to misleading conclusions.
		We also note that the correction to the OPE for the order and disorder fields that we identify in Eq.~\eqref{eq:smu_ope} is expected to be important for Ref.~\cite{VaeziFib}; see their Eq.~(14).
	}
%The symbols $\sigma$ and $\bar\sigma$ are ill-defined despite consistency with the $\Zth$ charges.%

(\textit{v}) Conversely---and perhaps more surprisingly---one can not generally back out lattice analogues of local fields (e.g., $\psi \bar\psi$) from knowledge of lattice realizations of chiral fields (e.g., $\psi$).
Equations~\eqref{Bexpansion} and \eqref{eq:OpPsi} again clearly illustrate the point.
The expansion of $(\hat\beta_{L,2a-1}+\hat\beta_{L,2a}) (\hat\beta_{R,2a-1}+\hat\beta_{R,2a}) \propto \hat\sigma_a(\omega\hat\tau_a + \omega^2\hat\tau_a^2 - 1)$ involves $s$---far more relevant than $\psi \bar\psi$---despite the fact that the leading terms in brackets individually yield $\psi$ and $\bar\psi$.
By expanding multiplicands,
\begin{align}
	(\hat\beta_{L,2a-1} + \hat\beta_{L,2a}) (\hat\beta_{R,2a-1} + \hat\beta_{R,2a})
	\sim (\psi + \partial\Phi_{\sigma\bar\epsilon} + \dots) (\bar\psi + \partial\Phi_{\epsilon\bar\sigma} + \dots) ,
\end{align}
we see that OPE's involving the \emph{subleading} terms can in fact generate terms more relevant than $\psi\bar\psi$.
(For example, $\partial\Phi_{\sigma\bar\epsilon} \times \bar\psi \rightarrow \Phi_{\sigma\bar\sigma} + \dots$.)
The actual lattice manifestation of $\psi\bar\psi$ is therefore much more complicated than this simple product, as seen from Eq.~\eqref{Bexpansion}.

\section{The order-disorder operator product expansion}
\label{sec:OPE}

The `dictionary' we sought to construct that relates lattice operators in the three-state Potts model to $\Zth$ CFT fields is now complete (at least for fields that are relevant or marginal at criticality).
There remains, however, one important result quoted earlier that we wish to now substantiate.
In particular, in Sec.~\ref{ParafermionsSection} we claimed that the parafermion fields $\psi$ and $\bar\psi$ do \emph{not} actually provide the dominant term in the OPE for the spin and disorder fields.
Here we will show using precise numerics that [as given in Eqs.~\eqref{eq:smu_ope}] the correct OPE instead reads
\begin{align} \label{smuOPE}
	s^\dag(z,\zbar) \mu(0,0) &= \frac{1}{(z\zbar)^{2/15}}
		\big[ C_{1} z^{1/15}\zbar^{2/5} \Phi_{\sigma\bar\epsilon}(0,0) + C_2 z^{2/3}\psi(0,0) + \dots \big] .
%\\	s^\dag(z,\zbar) \mu^\dag(0,0) &= \frac{1}{(z\zbar)^{2/15}}
%		\big[ C_1^\ast z^{2/5}\zbar^{1/15} \Phi_{\epsilon\bar\sigma}(0,0) + C_2^\ast \zbar^{2/3}\bar\psi(0,0) + \dots \big] .
\end{align}
(The OPE for $s^\dag$ and $\mu^\dag$ then follows from parity.)
We adopt the normalization convention that $\braket{\Phi_{f\bar{f}}(z,\zbar) \Phi_{f\bar{f}}^\dag(0,0)} = z^{-2h} \zbar^{-2\bar{h}}$ where the field $\Phi_{f\bar{f}}$ carries dimensions $(h,\bar{h})$.
With this convention, the second coefficient is known to be $C_2 = 1/\sqrt{3}$ \cite{FZ2}, while we conjecture the first to be exactly
\begin{align}
	|C_1| = \sqrt{\frac{\Gamma(\frac{1}{5})\Gamma(\frac{3}{5})^3} {2\Gamma(\frac{4}{5})\Gamma(\frac{2}{5})^3}} \approx 0.772 .
	\label{c1c2}
\end{align}
We discuss this conjecture further below but first focus on its implications.
Importantly, the above equation implies that $\Phi_{\sigma\bar\epsilon}$, which has scaling dimension $7/15<2/3$, constitutes the leading fusion product of $s^\dag$ and $\mu$.

We begin with a heuristic argument for Eq.~\eqref{smuOPE}.
Recall that the `left' lattice parafermion operators are defined by $\hat\beta_{L,2a-1}  = \omega^2 \hat\mu'_{a-\frac12} \hat\sigma^\dag_a$ and $\hat\beta_{L,2a}  = \hat\mu'_{a+\frac12} \hat\sigma^\dag_a$.
That is, up to factors of $\omega$ these operators differ in that the disorder string $\hat\mu'$ sits to the left of $\hat\sigma$ on odd parafermion sites but to the right of $\hat\sigma$ on even sites.
(As mentioned earlier the string $\hat\mu'_b$ is related to $\hat\mu_b$ by a conserved quantity: $\hat\mu'_b= \mathcal{Q}\hat\mu_b$.)
It is thus instructive to consider the analogous quantities in the continuum limit, where the spin field approaches the disorder field from the left or right (at equal times).
Eq.~\eqref{smuOPE} yields $s^\dag(\delta)\mu(0) = \kappa_1 \Phi_{\sigma\bar\epsilon}+\kappa_2\psi+\dots$, where $\delta$ is real, positive and small; $\kappa_{1,2}$ are constants.
To fuse the fields from the opposite direction we continue this equation to real negative argument; this gives $\omega^2 s^\dag(-\delta)\mu(0) = -\kappa_1 \Phi_{\sigma\bar\epsilon}+\kappa_2\psi+\dots$.
The important observation is that the operator product acquires an extra minus sign when fusing from the left versus right. This sign appears whether the continuation is clockwise or counterclockwise in the complex plane; the different continuations correspond only to different factors of $\omega$.
Taking symmetric and antisymmetric linear combinations $s^\dag(\delta)\mu(0) \pm \omega^2 s^\dag(-\delta)\mu(0)$ thus isolates either the parafermion field or $\Phi_{\sigma\bar\epsilon}$, which is entirely consistent with our previously established lattice operator/field correspondence given in Eqs.~\eqref{eq:betaLexpansion}.

Given the caveats discussed at the end of the previous section [particularly comment $(v)$], it is nevertheless worthwhile to supply direct numerical evidence of Eq.~\eqref{smuOPE}.
To this end we numerically compute 3-point correlators of $\hat\sigma_a^\dag$, $\hat\mu'_{1/2}$, and $\hat{b}^\pm_{x}$ on the lattice, where the last of these is defined as
\begin{align}
	\hat{b}^\pm_x \equiv \hat\beta_{L,2x}^\dag \pm \hat\beta_{L,2x-1}^\dag .
\end{align}
Equations~\eqref{eq:OpPsi} and \eqref{eq:OpSigmaE} tell us their field expansion:
\begin{subequations}\begin{align}
	\hat{b}^+ &= f_+ \psi^\dag + \dots ,
\\	\hat{b}^- &= f_- \Phi_{\sigma\bar\epsilon}^\dag + \dots ,
\end{align}\end{subequations}
with $f_+$ and $f_-$ constants.
%This is supported by Fig.~\ref{fig:CorrPlot_nonlocal}.
Conceptually speaking, one then expects the following equivalence between the lattice correlators and equal-time 3-point correlators in the $\Zth$ parafermion CFT,
\begin{subequations} \label{eq:smu_corr} \begin{align}
	\Braket{ \hat\sigma_a^\dag \; \hat\mu'_{1/2} \; \hat{b}^-_{x} }
		&\sim \Braket{ s^\dag(a-\tfrac12, a-\tfrac12) \; \mu(0, 0) \; \Phi_{\sigma\bar\epsilon}^\dag(x-\tfrac12, x-\tfrac12) }_\textrm{CFT} ,
	\label{eq:smu_corr_minus}
\\	\Braket{ \hat\sigma_a^\dag \; \hat\mu'_{1/2} \; \hat{b}^+_{x} }
		&\sim \Braket{ s^\dag(a-\tfrac12, a-\tfrac12) \; \mu(0, 0) \; \psi^\dag(x-\tfrac12, x-\tfrac12) }_\textrm{CFT} .
	\label{eq:smu_corr_plus}
\end{align} \end{subequations}
Thus evaluation of the first correlator allows us to extract the coefficient $C_1$ and demonstrate the presence of $\Phi_{\sigma\bar\epsilon}$ in the OPE above.

For a generic lattice operator $\hat{O}$ at the 0\textsuperscript{th} site, evaluating $\braket{\hat{O}_0 \hat{b}^-_x}$ as a function of $x$ allows us to extract the $\Phi_{\sigma\bar\epsilon}$ piece within $\hat{O}$. Here $\hat{O}$ must carry charges $Q_3 = 2$ and $Q_3^\textrm{dual} = 1$ for the expectation value to be non-vanishing.
Thus if $\hat{O} = c_{\sigma\bar\epsilon}\Phi_{\sigma\bar\epsilon}+{\rm other~terms}$, then $\braket{\hat{O}_0 \hat{b}^-_x} = c_{\sigma\bar\epsilon} f_- x^{-14/15} + \dots$ for asymptotically large $x$.
The subleading contribution comes from components of $\partial_x \Phi_{\sigma\bar\epsilon}$ within $\hat{O}$ and decays as $x^{-29/15}$.%
	\footnote{Note that components of $\psi$ in $\hat{O}$ would yield a power law $x^{-7/3}$ in the correlator, as $\hat{b}^-$ contains $\partial_x \psi^\dag$ but not $\psi^\dag$. This would generate the second subleading correction to the $x^{-14/15}$ scaling.}
A similar logic would allow us to extract the component of $\psi$ within $\hat{O}$:
if $\hat{O} = c_{\psi}\psi +{\rm other~terms}$, then $\braket{\hat{O}_0 \hat{b}^+_x} = c_{\psi} f_+ x^{-4/3} + \dots$\ .
Here the subleading part scales as $x^{-29/15}$ and arises from the $\Phi_{\sigma\bar\epsilon}$ and $\partial_x \Phi_{\sigma\bar\epsilon}^\dag$ pieces in $\hat{O}$ and $\hat{b}^+$, respectively.

\begin{figure}[t]
	\centering
	\subfigure[ $\big|\!\braket{\hat\sigma_a^\dag \hat\mu'_{1/2} \hat{b}^-_x}\!\big|$ vs.~$x$ ]{ \includegraphics[width=67mm]{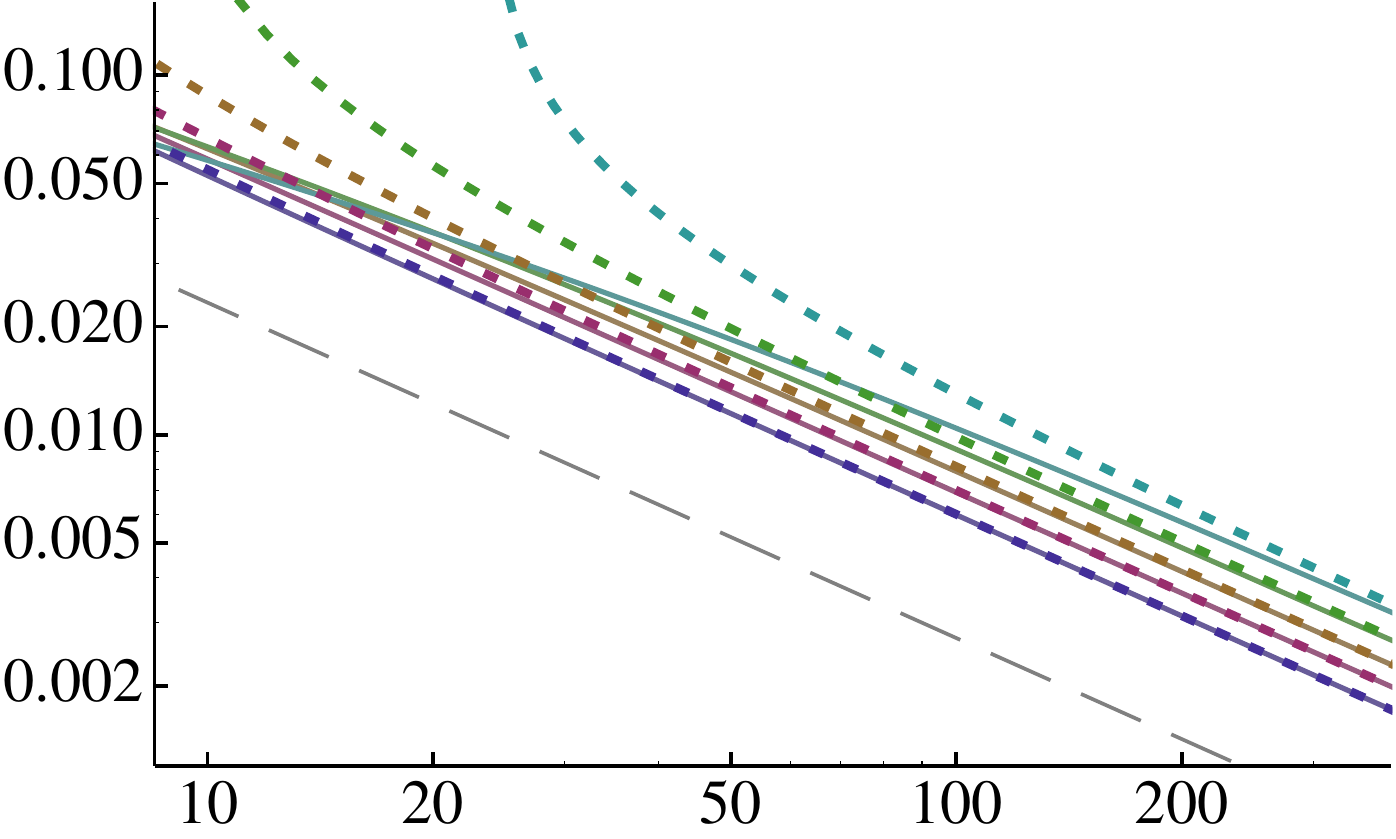} \label{fig:smu_bminus} }
		\;
	\subfigure[ $\big|\!\braket{\hat\sigma_a^\dag \hat\mu'_{1/2} \hat{b}^+_x}\!\big|$ vs.~$x$ ]{ \includegraphics[width=67mm]{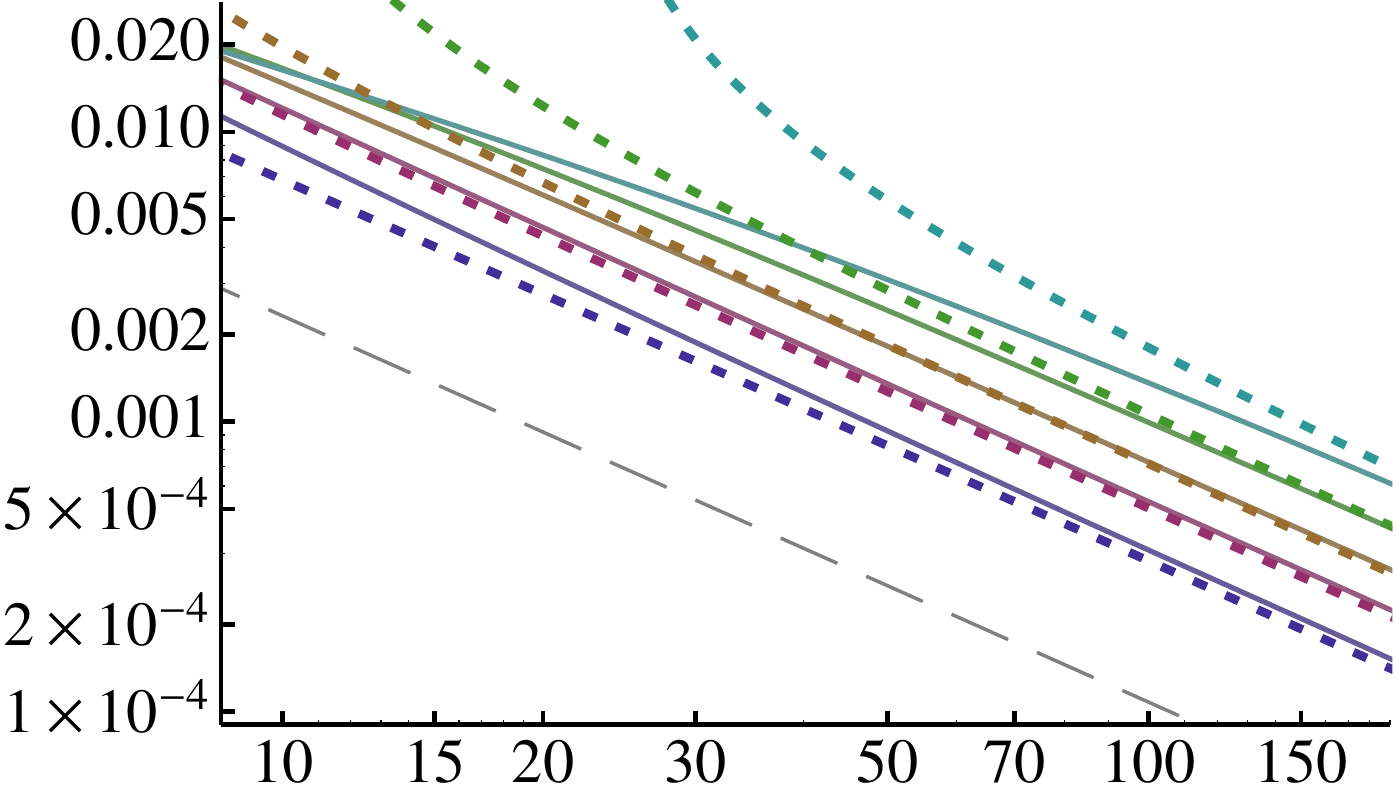}  \label{fig:smu_bplus} }
	\includegraphics[width=16mm]{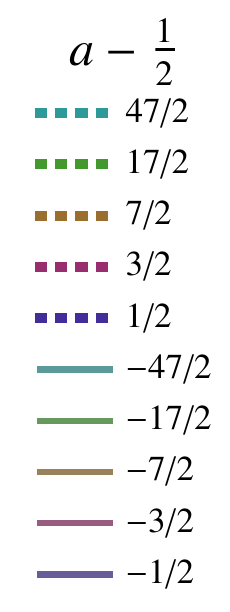}
		\\
	\subfigure[ $\displaystyle\lim_{x\to\infty} \big|\!\braket{\hat\sigma_a^\dag \hat\mu'_{1/2} \hat{b}^-_x}\!\big|x^{14/15}$ vs.~$a-\frac12$ ]{ \includegraphics[width=67mm]{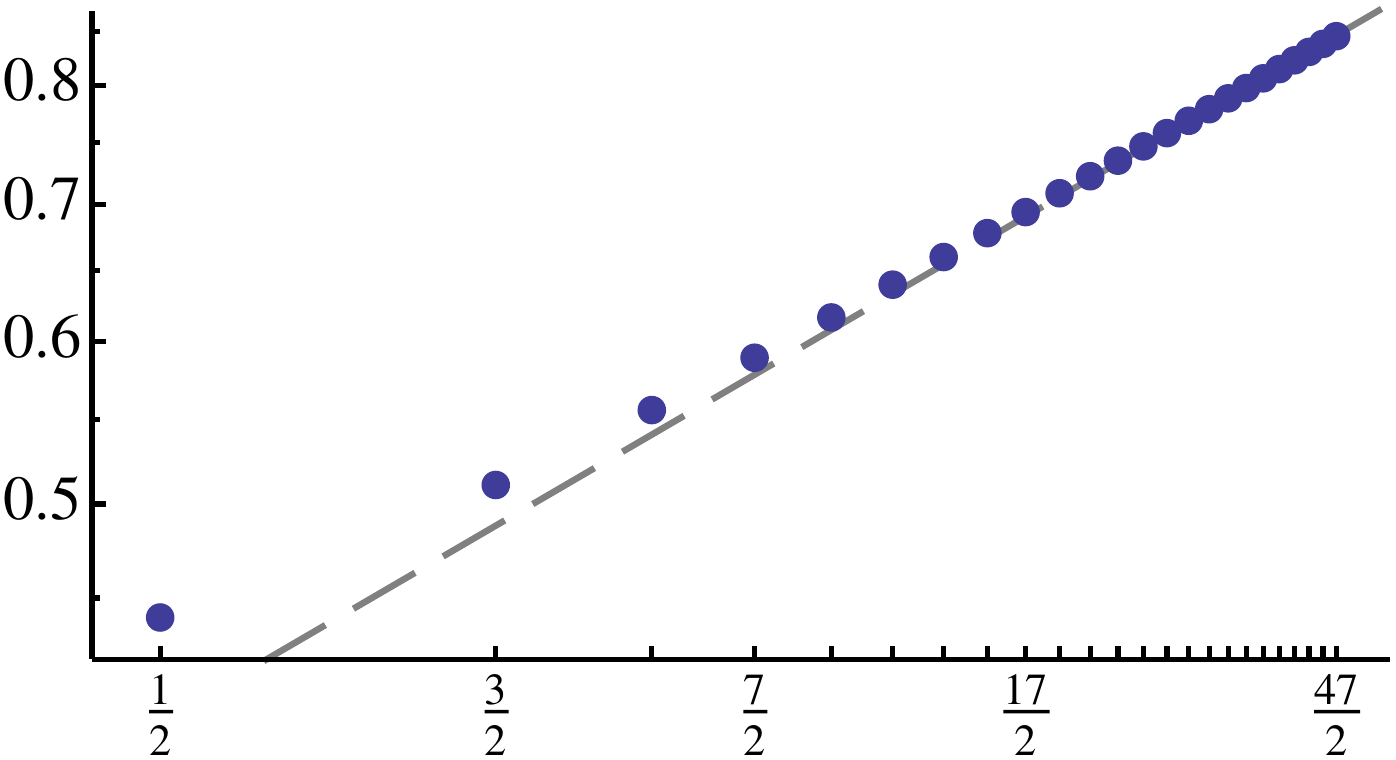} \label{fig:smu_bminuscoef} }
		\qquad
	\subfigure[ $\displaystyle\lim_{x\to\infty} \big|\!\braket{\hat\sigma_a^\dag \hat\mu'_{1/2} \hat{b}^+_x}\!\big|x^{4/3}$ vs.~$a-\frac12$ ]{ \includegraphics[width=67mm]{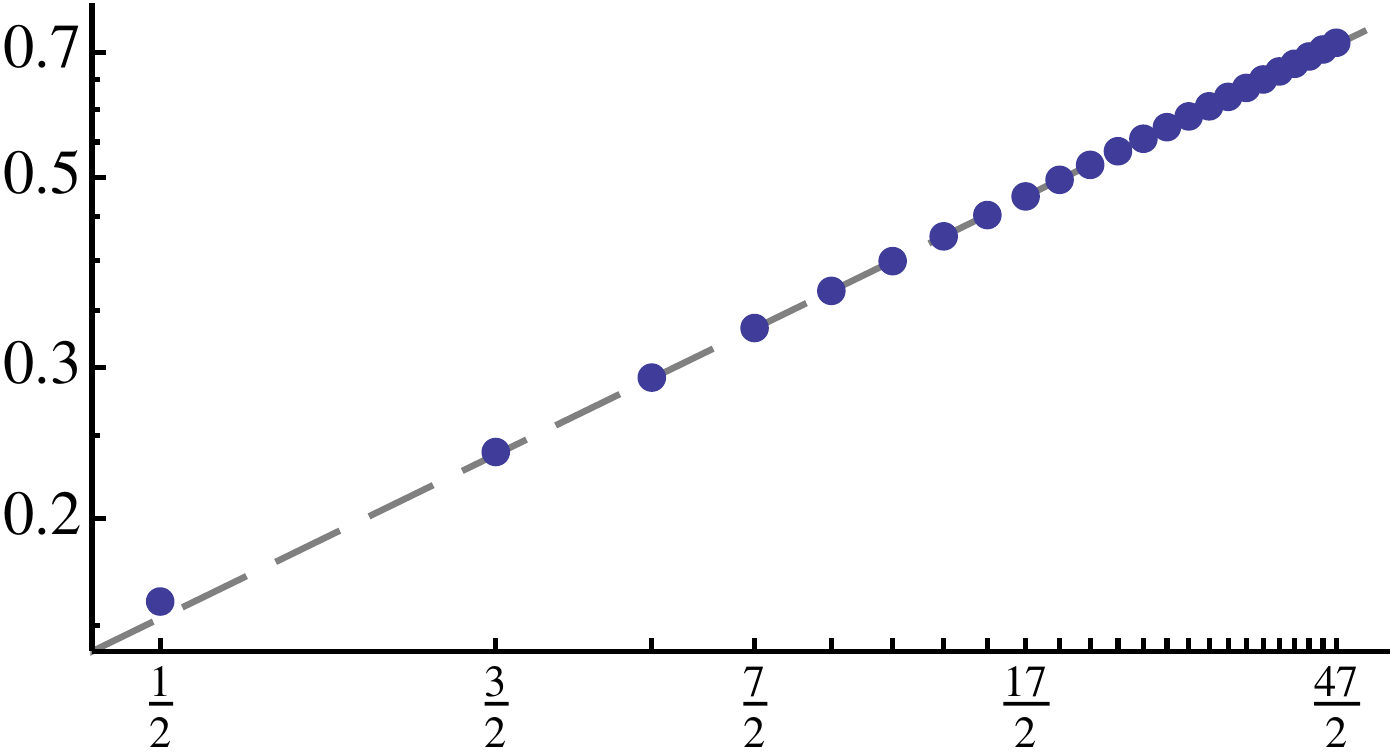}  \label{fig:smu_bpluscoef} }
	\caption{%
		(Top) Absolute value of the correlation functions $\braket{\hat\sigma_a^\dag \hat\mu'_{1/2} \hat{b}^\pm_x}$ versus $x$ for the set of fixed $a$ listed on the right.
			%The solid lines and dotted lines denote $a<\frac12$ and $a>\frac12$ respectively; from smallest to largest, $|a-\frac12| = \frac{1}{2}, \frac{3}{2}, \frac{7}{2}, \frac{17}{2}, \frac{47}{2}$.
			As a guide to the eye, the dashed gray line is proportional to $x^{-14/15}$ for (a) and $x^{-4/3}$ for (b).
			The data show that the respective correlators approach these power laws for large $x$.
		(Bottom)
			Coefficients for fields (c) $\Phi_{\sigma\bar\epsilon}$ and (d) $\psi$ within the lattice operator $\hat\sigma_a^\dag \hat\mu'_{1/2}$, ploted as a function of $a - \frac12$.
			The data show excellent agreement with the scaling form $|a-\frac12|^{1/5}$ for (c) and $|a-\frac12|^{2/5}$ for (d).  Together, these results confirm the modified OPE for the spin and disorder field in Eq.~\eqref{smuOPE}.
	}
\end{figure}

We numerically computed the left-hand sides of Eqs.~\eqref{eq:smu_corr} via DMRG; Figs.~\ref{fig:smu_bminus} and \ref{fig:smu_bplus} present the results.
The data show that for large $x$ and fixed $a$, the correlator $\big|\!\braket{\hat\sigma_a^\dag \hat\mu'_{1/2} \hat{b}^-_x}\!\big|$ decays as $x^{-14/15}$,
	while $\big|\!\braket{\hat\sigma_a^\dag \hat\mu'_{1/2} \hat{b}^+_x}\!\big|$ decays as $x^{-4/3}$.
In light of the preceding discussion we can thus conclude that the lattice operator $\hat\sigma_a^\dag \hat\mu'_{1/2}$ contains both $\psi$ \emph{and} $\Phi_{\sigma\bar\epsilon}$ field components.
This fact alone does not imply anything about the OPE of the fields $s^\dag$ and $\mu$, however, as the presence of $\Phi_{\sigma\bar\epsilon}$ may arise from OPE's between the less relevant terms within $\hat\sigma^\dag$ and $\hat\mu'$; see comment $(v)$ from Sec.~\ref{sec:chiral}.
Rather, we must additionally examine the behavior of the correlators as a function of $a$ to determine the structure of the OPE.

We therefore extract from our data the $a$-dependent coefficient for the leading term in each lattice correlator,
\begin{subequations} \begin{align}
	\Braket{ \hat\sigma_a^\dag \; \hat\mu'_{1/2} \; \hat{b}^-_{x} }
		&= c_{\sigma\bar\epsilon}(a) x^{-14/15} + \mathcal{O}(x^{-29/15}) ,
\\	\Braket{ \hat\sigma_a^\dag \; \hat\mu'_{1/2} \; \hat{b}^+_{x} }
		&= c_\psi(a) x^{-4/3} + \mathcal{O}(x^{-29/15}).
\end{align} \end{subequations}
Figures \ref{fig:smu_bminuscoef} and \ref{fig:smu_bpluscoef} plot the results.
Recall that the coefficient $c_{\sigma\bar\epsilon}(a)$ captures the amount of $\Phi_{\sigma\bar\epsilon}$ found in the operator $\hat\sigma_a^\dag \hat\mu'_{1/2}$.
If $\Phi_{\sigma\bar\epsilon}$ indeed appears in the OPE of $s^\dag(z,\zbar)$ and $\mu(0,0)$, then its coefficient scales as $|z|$ to the power $\operatorname{dim}(\Phi_{\sigma\bar\epsilon}) - \operatorname{dim}(s) - \operatorname{dim}(\mu) = \frac{7}{15} - \frac{2}{15} - \frac{2}{15} = \frac{1}{5}$.
If instead $\Phi_{\sigma\bar\epsilon}$ comes from an OPE of less relevant fields---say $\Phi_A(z,\zbar)$ and $\Phi_B(0)$---then one would expect a smaller/negative power since by definition $\operatorname{dim}(\Phi_A) + \operatorname{dim}(\Phi_B) > \frac{2}{15} + \frac{2}{15}$.
Similar logic allows one to deduce the scaling form of $c_\psi(a)$, which again is the coefficient of $\psi$ within $\hat\sigma_a^\dag \hat\mu'_{1/2}$.
Assuming that the modified OPE in Eq.~\eqref{smuOPE} holds, we therefore obtain the following scaling ansatz for the coefficients,
\begin{subequations} \begin{align}
	\big|c_{\sigma\bar\epsilon}(a)\big| \propto C_1 \big| a-\tfrac12 \big|^{1/5} + \mathcal{O}\left( \big|a-\tfrac12\big|^{-1} \right) ,
\\	\big|c_\psi(a)\big| \propto C_2 \big| a-\tfrac12 \big|^{2/5} + \mathcal{O}\left( \big|a-\tfrac12\big|^{-8/5} \right) .
\end{align} \end{subequations}
Figure~\ref{fig:smu_bminuscoef} shows that $c_{\sigma\bar\epsilon}(a)$ indeed scales as $\big| a-\tfrac12 \big|^{1/5}$.  This confirms our hypothesis that the OPE between the spin field and disorder field conforms to Eq.~\eqref{smuOPE}.  [Figure \ref{fig:smu_bpluscoef} is also consistent with the scaling above, but this merely reaffirms the existence of $\psi$ within the OPE.]

Interestingly, we can further extract from our numerics the coefficients $C_1$ and $C_2$ appearing in the OPE.
%Here we adopt the normalization convention that $\braket{\Phi_{f\bar{f}}(z,\zbar) \Phi_{f\bar{f}}^\dag(0,0)} = z^{-2h} \zbar^{-2\bar{h}}$ where $\Phi_{f\bar{f}}$ have scaling dimensions $(h,\bar{h})$.
Via fits to the 2-point correlators of $\hat\sigma$, $\hat\mu$, and $\hat{b}^\pm$ operators, we find the scale factors
\begin{align}\begin{split}
	\hat\sigma_x &\approx 0.7839 \, s(x) + \dots ,
\\	\hat\mu_x    &\approx 0.7839 \, \mu(x) + \dots ,
\\	\hat{b}^+_x  &\approx 0.5649 \, \psi^\dag(x) + \dots ,
\\	\hat{b}^-_x  &\approx 0.9369 \, \Phi_{\sigma\bar\epsilon}^\dag(x) + \dots .
\end{split}\end{align}
Moreover, fitting to the 3-point correlator data shown in Figs.~\ref{fig:smu_bminuscoef} and \ref{fig:smu_bpluscoef} yields $|c_{\sigma\bar\epsilon}(a)| \approx 0.45 |a-\frac12|^{1/5}$ and $|c_\psi(a)| \approx 0.20 |a-\frac12|^{2/5}$.
The scale factors above then allow us to back out $|C_2| \approx 0.58$, in good agreement with the exact value $C_2 = 1/\sqrt{3} \approx 0.577$, together with $|C_1| \approx 0.78$.
The latter is consistent with Eq.~\eqref{c1c2}.

%where $C_1$ is computed via the Coulomb-gas formalism for the tetracritical Ising CFT while $C_2$ is given in Ref.~\cite{FZ2}.  \jason{Are we referencing someone else's work for $C_1$?}
%The moduli of the OPE coefficients are given by $|C_1|^2 = \frac{\Gamma(\frac{1}{5})\Gamma(\frac{3}{5})^3} {2\Gamma(\frac{4}{5})\Gamma(\frac{2}{5})^3}$ and $C_2^2 = \frac{1}{3}$.
To motivate our conjecture for $C_1$, we use insights from the tetracritical Ising CFT. Because the two CFT's are related by orbifolding, their respective operators and their products are also related.
We denote the primary fields in the tetracritical Ising model $\phi_{p,q}$, where $p$ and $q$ are Kac labels---integers obeying $1\le p \le 4$, $1\le q\le 5$, and $p + q \equiv 0 \pmod 2$.
The fields $s$ and $\mu$ in the Potts CFT are related to $\phi_{3,3}$, all having the same scaling dimension 2/15.
The OPE of $\phi_{3,3}$ with itself contains all the fields $\phi_{1,3}$, $\phi_{1,5}$, $\phi_{3,1}$, $\phi_{3,3}$, and $\phi_{3,5}$.
The first three correspond  to $1$, $\psi$, and $W$, while the latter three map to $X$, $\sigma$, and $\epsilon$, respectively.
Thus it is reasonable to expect that the OPE in the Potts CFT between $s$ and $\mu$ consists of both $\psi$ and $\epsilon$ fields, and that the coefficients are also related to those of the tetracritical Ising model.
Indeed, our conjecture can alternatively be phrased as $|C_1| = \big[\sqrt{2}\, C^{(3,3)}_{(3,3),(3,3)} C^{(3,5)}_{(3,3),(3,3)}\big]^{1/2}$, where $C^{(p'',q'')}_{(p,q),(p',q')}$ are structure coefficients for the tetracritical Ising CFT given in Ref.~\cite{DotsenkoFateev}.
This could presumably be derived explicitly by working out four-point functions following Ref.~\cite{DiFrancesco}; however, we believe our numerical check is convincing.

%%%%%%%%%%%%%%%%%%%%%%%%%%%%%%%%%%%%%%%%%%%%%%%%%%%%%%%%%%%%%%%%%%%%%%%%%%%%%%%
%%%%%%%%%%%%%%%%%%%%%%%%%%%%%%%%%%%%%%%%%%%%%%%%%%%%%%%%%%%%%%%%%%%%%%%%%%%%%%%
\section{Conclusions}

%\roger{Topological protection, based on the argument from the PRX, but I don't know how to show this rigourously}
%
%This implies that the two ground states cannot be distinguished by parafermion operators/combinations of parafermion operators, such as
%$\psi(x)$, $\psi(x)\bar\psi(x')$, etc.
%In addition, the Fibonacci spectra must be robust to small deformations of the Hamiltonian with parafermion fields, for example
%\begin{align}
%	H = H_\textrm{CFT} - \int_{x,x'}\!\lambda(x-x') \psi(x) \bar\psi(x') + \hc
%\end{align}
%
%Since the entire sequence of Read-Rezayi states are known to exist, much of this should also apply to $\mathbb{Z}_N$ critical parafermion theories.
Over the course of this paper we illustrated how one can make a remarkable amount of progress in connecting fields in a $\Zth$ parafermion conformal field theory to microscopic lattice operators in the three-state Potts model.
Indeed, we constructed lattice analogues of all local fields with scaling dimensions $\leq 2$, only some of which were previously known.
This includes the \emph{individual} components $T$ and $\bar T$ of the energy-momentum tensor; although their sum is of course given by the Hamiltonian, the lattice analogue of their difference is far from obvious.
We believe this represents the first example of a lattice analogue for these components in an interacting unitary field theory (although there have been some interesting developments utilizing Schramm-Loewner Evolution\cite{DoyonRivaCardySLE}).
Just as interesting, we also found lattice realizations of the relevant \emph{chiral} primary fields including the physically important parafermions.%
	\footnote{The one relevant physical field for which we did not construct a lattice analogue is $\Phi_{\sigma \Xbar}$.  This is possibly available from R.\ Mong upon request.}
By Hermitian conjugation and/or dualizing our lattice operators we can in fact construct analogues for all 18 primaries!
Outside of simple free theories like the Ising model, obtaining such correspondences is notoriously difficult---especially for chiral fields since by definition their lattice counterparts must be non-local.

One important application of these results is that they allow one to controllably access exotic two-dimensional topologically ordered phases starting from weakly coupled chains.
Suppose that in the decoupled-chain limit a system is characterized by a Hamiltonian $H_0 = \sum_y H^\textrm{crit}_y$, where $H^\textrm{crit}_y$ is the Hamiltonian for a critical chain $y$ of lattice parafermion operators of the type we have studied here.
The low-energy physics of the chains is then well-known---each one is described by a $\Zth$ parafermion conformal field theory.  Understanding the fate of the system in the presence of weak interchain perturbations $\delta H$ generally poses a highly nontrivial problem due to the absence of a free-particle description.  However, the ultraviolet-infrared correspondences we have derived allow one to quite generally expand $\delta H$ in terms of continuum fields, thereby `filtering out' the unimportant high-energy physics.  This approach was used in Ref.~\cite{Mong}---where Eqs.~\eqref{eq:betaRexpansion} and \eqref{eq:betaLexpansion} were first quoted---along with Ref.~\cite{BarkeshliJiang} to construct phases supporting Fibonacci anyons.  Without this filtering procedure the physics would be much more obscure, presumably requiring sophisticated numerics to sort out.  Interestingly, it is even possible to study from the lattice point of view the perturbed conformal field theories giving rise to this `Fibonacci phase'.  Viewing the problem from this lens gives great insight into the universal topological properties of the two-dimensional system, but in the framework of trivially solvable one-dimensional models \cite{Aasen}.

%A related application is in understanding off-critical systems obtained by perturbing conformal field theories. This is a crucial part of the construction of Fibonacci anyons described above, but is a problem of much more general interest. \paul{not sure really what the connection is here, other than Potts!}

While a completely general method for mirroring strongly interacting field theories on the lattice remains unclear, our approach does suggest a strategy for profitably attacking other nontrivial examples.
It is worth emphasizing, for instance, how far symmetry took us in establishing our lattice operator/CFT field dictionary.
The basic requirement was a link between the spin and disorder fields $s, \mu$ to the lattice spin and disorder operators $\hat\sigma_a, \hat\mu_b$.
The identification $\hat\sigma_a \sim s$ and $\hat\mu_b\sim \mu$  at criticality is easily guessed but could instead have been uniquely inferred by computing two-point lattice correlators numerically.
This simple correspondence then allowed us to transcribe how microscopic symmetries of the quantum Potts chain transform not just $s$ and $\mu$, but in fact \emph{all} of the physical primary fields (one can infer the symmetry properties of the others through consistency with operator product expansions).
Much of our dictionary---including the chiral operators---can be assembled by carefully matching the quantum numbers carried by the lattice and continuum fields.  We also note that symmetry considerations strongly suggest the correction to the operator product expansion for $s,\mu$ that we introduced, though numerics was ultimately necessary to make a compelling case.

In some cases, however, symmetry alone proves insufficient for identifying lattice analogues of continuum fields.
Suppose that two fields $\mathcal{F}_1$ and $\mathcal{F}_2$ both transform in the same way under all symmetries as a pair of lattice operators $\hat O$ and $\hat O'$; moreover, let $\mathcal{F}_2$ be less relevant than $\mathcal{F}_1$.
Symmetry then predicts the generic expansions $\hat O \sim a_1 \mathcal{F}_1 + a_2 \mathcal{F}_2+\dots$ and $\hat O' \sim a'_1 \mathcal{F}_1 + a'_2 \mathcal{F}_2+\dots$ only up to (unknown) non-universal coefficients $a_{1,2}$ and $a'_{1,2}$.
Finding the precise linear combination of lattice operators that kills the $\mathcal{F}_1$ bit, leaving the less-relevant $\mathcal{F}_2$ as the leading piece, requires additional input.
Such a scenario arose here in the context of the parafermion bilinear $\psi\bar\psi$, which transforms identically to the more relevant spin field.
In this case integrability provided the tool necessary for constructing a lattice equivalent of $\psi\bar \psi$.
A similar challenge arose for the stress-energy tensor; there we supplemented symmetry with numerics to back out lattice analogues of the individual components $T$ and $\bar{T}$.

The above methodology appears rather general.
It is plausible that symmetry combined with numerics and integrability provides a sufficient toolkit for bridging any nontrivial conformal field theory with an appropriate lattice model.
For simplicity we have focused on quantum chains, but these methods also should allow the determination of field analogues in the corresponding two-dimensional classical systems.
This would allow contact with results on operators representing various non-local geometric observables in the Potts models \cite{Vasseur}.
%(Though if a large number of relevant fields transform in the same way, isolating each piece numerically could be quite cumbersome.)
Potentially these results may prove useful as well in finding operators satisfying the full set of lattice Cauchy-Riemann equations.

%%%%%%%%%%%%%%%%%%%%%%%%%%%%%%%%%%%%%%%%%%%%%%%%%%%%%%%%%%%%%%%%%%%%%%%%%%%%%%%
\section*{Acknowledgment}

It is a pleasure to thank Erez Berg, Chetan Nayak, Miles Stoudenmire, and Mike Zaletel for helpful discussions related to this work.
We also acknowledge funding from the NSF through grants DMR-1341822 (D.~C.\ \& J.~A.) and DMR/MPS1006549 (P.~F.);
the Sherman Fairchild Foundation (R.~M.);
the Alfred P.~Sloan Foundation (J.~A.);
the Bi-National Science Foundation and I-Core:
the Israel Excellence Center ``Circle of Light'' (N.~L.);
the Caltech Institute for Quantum Information and Matter, an NSF Physics Frontiers Center with support of the Gordon and Betty Moore Foundation;
and the Walter Burke Institute for Theoretical Physics at Caltech.
This work was completed at the Topological Phases and Quantum Computation Workshop 2014 under the hospitality of the Moorea Ecostation Center for Advanced Studies.

\appendix
%%%%%%%%%%%%%%%%%%%%%%%%%%%%%%%%%%%%%%%%%%%%%%%%%%%%%%%%%%%%%%%%%%%%%%%%%%%%%%%

\comment{\section{Duality acting on parafermion operators}
\begin{align}\begin{split}
		\hat\sigma_{a+1}^\dag \hat\mu_a^\dag
		&=	\dual^2[ \hat\sigma_{a}^\dag \hat\mu_{a-1}^\dag ]
	=	\dual[ e^{i\theta} \hat\sigma_a^\dag \hat\mu_a^\dag ]
       = \dual[ e^{i\theta} \hat\sigma_a^\dag \hat\mu_{a-1}^\dag \hat\tau_{a}^\dag ]
	= e^{i\theta} \dual[ \hat\sigma_a^\dag \hat\mu_{a-1}^\dag ] \dual[ \hat\tau_{a}^\dag ]
	= e^{i\theta} (e^{i\theta} \hat\sigma_a^\dag \hat\mu_{a}^\dag) (\hat\sigma_{a+1}^\dag \hat\sigma_a^\phd) \\	
&= e^{2i\theta} \omega \hat\sigma_{a+1}^\dag \hat\mu_{a}^\dag\ .
\end{split}\end{align}
Hence $e^{2i\theta} \omega = 1$.  Requiring that $(\dual[\hat\beta_{R,2a-1}])^3 = 1$
yields $e^{i\theta} = \omega$, so that $$\dual[\hat\beta_{R,2a-1}] = \hat\beta_{R,2a}\ .$$ A similar argument using the alternative dual transformation $\dual'$ shows that $\dual'[\hat\beta_{L,b}] = \hat\beta_{L,b-1}$.
\paul{what about others, e.g. $\dual$ on $\beta_L$ ?}

Using these, we determine the phase $e^{i\theta}$ in $\dual[ \hat\sigma^\dag_a \hat\mu^\dag_{a-1} ] = e^{i\theta} \hat\sigma^\dag_a \hat\mu^\dag_a$\ :
}

\section{DMRG numerical methods}

Numerical computations were done via the density-matrix-renormalization-group (DMRG) method \cite{WhiteDMRG, SchollwockDMRG, SchollwockDMRGMPS, McCullochDMRGMPS, McCullochDMRG}.
In particular, we use the two-site update infinite DMRG algorithm described in Ref.~\cite{Kjall}, with the ``bond dimension'' $\chi$ as the control parameter.
All numerical results presented in this paper require careful extrapolation of the data as $\chi \to \infty$.

A faithful representation for the ground state at criticality would require an infinite bond dimension, and thus any simulation at finite $\chi$ is an approximation to the critical state.
Nevertheless, one can perform ``finite entanglement scaling'' \cite{Tagliacozzo08,PollmannScaling} to extrapolate the system's thermodynamic behavior.
The measure for how closely one approaches criticality is given by the correlation length $\xi$ of the ground-state wavefunction computed by the transfer matrix technique.
While $\xi$ must be finite for any finite dimension $\chi$, it grows rapidly as $\chi$ is increased.
One can intepret $\xi$ as the ``effective system size'' for the DMRG simulation, and thus finite entanglement scaling is analogous to finite size scaling for exact diagonalization/Monte Carlo techniques.
We note that the correlation length $\xi$ scales algebraically with $\chi$, as do the computation cost for DMRG; therefore the computation resources for finite entanglement scaling is a polynomial function of the effective system size.

\begin{figure}[t]
	\centering
	\subfigure[]{ \includegraphics[width=67mm]{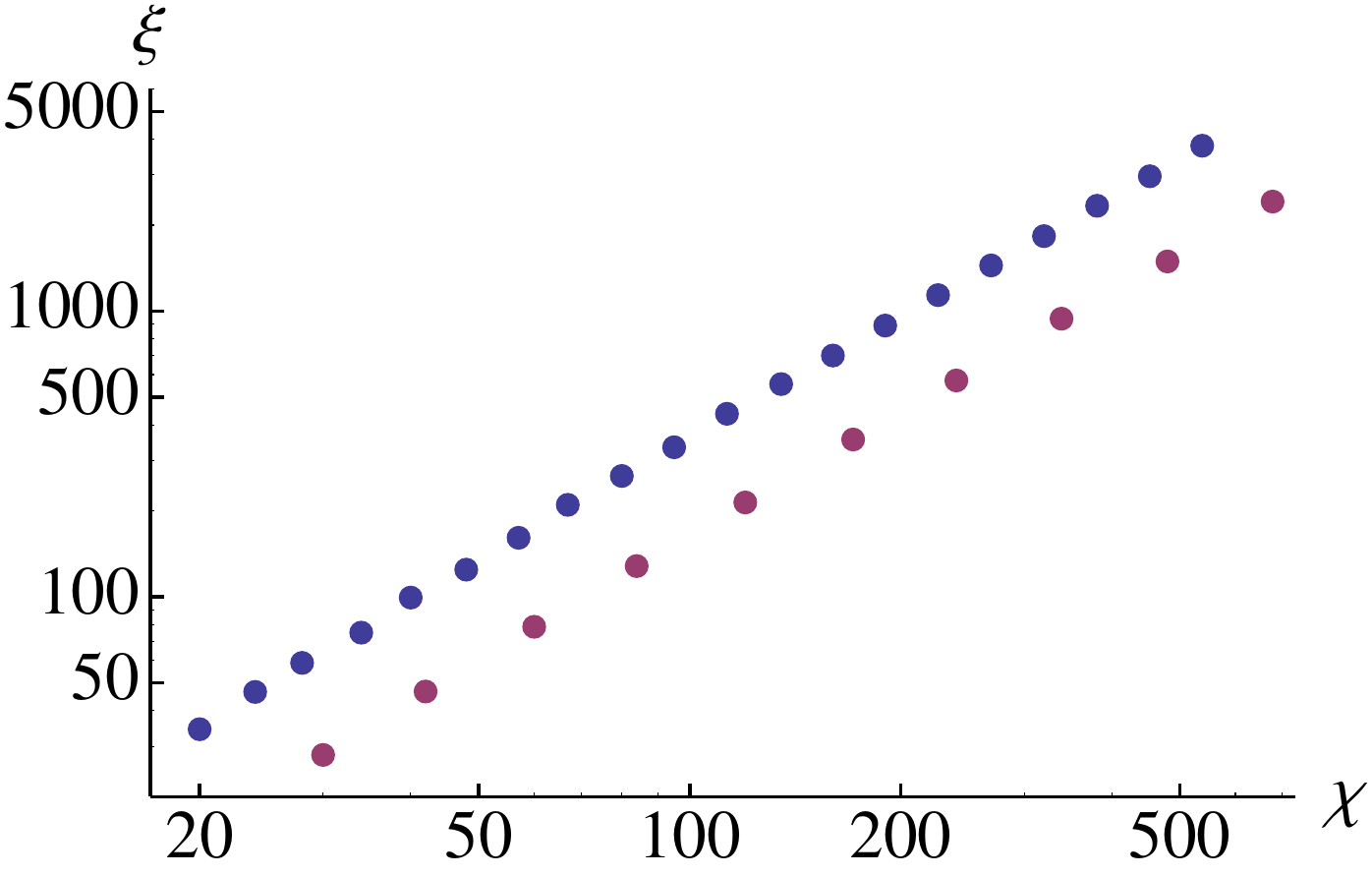} }
	\subfigure[]{ \includegraphics[width=67mm]{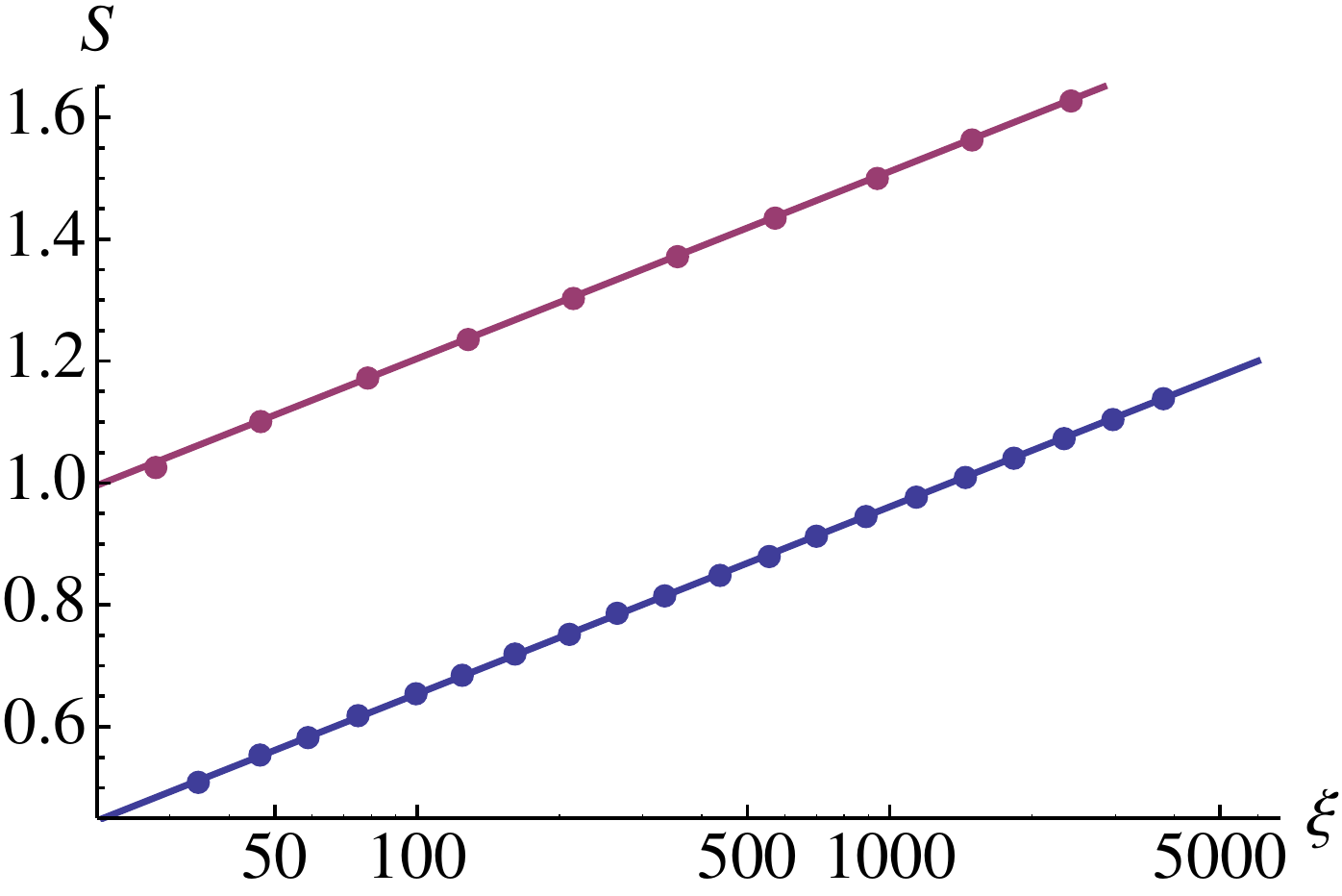} }
	\caption{%
		Demonstration of ``finite entanglement'' scaling from DMRG simulations where $\Zth$ symmetry is enforced (purple) and not conserved (blue).
		(a) Correlation length $\xi$ vs.\ the bond dimension $\chi$.
			The data show that the effective system size grows with rapidly bond dimension.
		(b) Entanglement entropy $S$ vs.\ $\xi$ at various bond dimensions.
			For both data sets, the entanglement entropy is given precisely by $S = \frac{c}{6}\log\xi + \textrm{const}$ for central charge $c = 4/5$.
	}
	\label{fig:DMRG}
\end{figure}

Figure~\ref{fig:DMRG} shows the scaling of $\xi$ as a funtion of $\chi$.
The purple data set was taken while enforcing $\Zth$ symmetry of the ground state, while the blue data set allows for a state with spontaneously broken symmetry.
Both sets of simulations approach criticality with increasing $\chi$, but from the ferromagnetic/paramagnetic directions for the blue/purple data.
This provides an effective way to extract observables (i.e., correlation functions) at criticality.

%%%%%%%%%%%%%%%%%%%%%%%%%%%%%%%%%%%%%%%%%%%%%%%%%%%%%%%%%%%%%%%%%%%%%%%%%%%%%%%%

\end{document}